\documentclass[longauth]{aa}  

\newcommand*{\mstar}{\ensuremath{M_\star\ }}
\newcommand*{\msun}{\ensuremath{M_\odot\ }}

\newcommand*{\fesc}{\ensuremath{f_{esc}\ }}

\newcommand*{\Ha}{H\ensuremath{\alpha}\ }
\newcommand*{\Hb}{H\ensuremath{\beta}\ }
\newcommand*{\Hg}{H\ensuremath{\gamma}\ }

\newcommand*{\xOIIItwo}{[O\,{\scshape ii}]\,\ensuremath{\lambda4959 \AA}}

\newcommand*{\xOII}{[O\,{\scshape ii}]\,\ensuremath{\lambda3727 \AA }}
\newcommand*{\xOIII}{[O\,{\scshape iii}]\,\ensuremath{\lambda5007 \AA}}

\newcommand*{\xNII}{[N\,{\scshape ii}]\,\ensuremath{\lambda6583 \AA}}

\newcommand*{\sigmaSFR}{\ensuremath{\Sigma_{\rm SFR}\ }}

\newcommand*{\OIII}{[O\,{\scshape iii}]}
\newcommand*{\OII}{[O\,{\scshape ii}]}

\newcommand*{\NII}{[N\,{\scshape ii}]}

\usepackage{graphicx}

\usepackage[colorlinks=true,
            linkcolor=red,
            urlcolor=red,
            citecolor=blue]{hyperref}

\usepackage{txfonts}

\newcommand{\myemail}{antonello.calabro@inaf.it}

\usepackage{multirow,array}

\usepackage{tikz}

\definecolor{lime}{HTML}{A6CE39}
\DeclareRobustCommand{\orcidicon}{%
    \begin{tikzpicture}
    \draw[lime, fill=lime] (0,0) 
    circle [radius=0.16] 
    node[white] {{\fontfamily{qag}\selectfont \tiny ID}};
    \draw[white, fill=white] (-0.0625,0.095) 
    circle [radius=0.007];
    \end{tikzpicture}
    \hspace{-2mm}
}

\foreach \x in {A, ..., Z}{%
    \expandafter\xdef\csname orcid\x\endcsname{\noexpand\href{https://orcid.org/\csname orcidauthor\x\endcsname}{\noexpand\orcidicon}}
}

\begin{document}



\title{The evolution of the SFR and $\Sigma_{SFR}$ of galaxies in cosmic morning ($4<z<10$)} 




\author{
A.Calabr{\`o}\inst{1}
\and L.Pentericci\inst{1}
\and P.Santini\inst{1}
\and A.Ferrara\inst{2}
\and M.Llerena\inst{1} 
\and S.Mascia\inst{1}
\and L.Napolitano\inst{1}
\and L.Y.A.Yung\inst{3}
\and L.Bisigello\inst{4,5}
\and M.Castellano\inst{1}
\and N.J.Cleri\inst{6,7}
\and A.Dekel\inst{8}
\and M.Dickinson\inst{9}
\and M.Franco\inst{10}
\and M.Giavalisco\inst{11}
\and M.Hirschmann\inst{12}
\and B.W.Holwerda\inst{13}
\and A.M.Koekemoer\inst{3}
\and R.A.Lucas\inst{3}
\and F.Pacucci\inst{14,15}
\and N.Pirzkal\inst{16}
\and G.Roberts-Borsani\inst{17}
\and L.M.Seill\'e\inst{18}
\and S.Tacchella\inst{19,20}
\and S.Wilkins\inst{21,22}
\and R.Amor\'in\inst{23,24}
\and P.Arrabal Haro\inst{9}
\and M.B.Bagley\inst{10}
\and S.L.Finkelstein\inst{10}
\and J.S.Kartaltepe\inst{25}
\and C.Papovich\inst{7,26}
}

\institute{INAF - Osservatorio Astronomico di Roma, via Frascati 33, 00078, Monte Porzio Catone, Italy (\myemail) 
\and Scuola Normale Superiore, Piazza dei Cavalieri 7, 50126 Pisa, Italy 
\and Space Telescope Science Institute, 3700 San Martin Drive, Baltimore, MD 21218, USA 
\and INAF - Istituto di Radioastronomia, via Piero Gobetti 101, 40129 Bologna, Italy 
\and Dipartimento di Fisica e Astronomia ``G.Galilei'', Universit\'a di Padova, Via Marzolo 8, I-35131 Padova, Italy 
\and Department of Physics and Astronomy, Texas A\&M University, College Station, TX, 77843-4242 USA 
\and George P.\ and Cynthia Woods Mitchell Institute for Fundamental Physics and Astronomy, Texas A\& M University, College Station, TX, 77843-4242 USA 
\and Racah Institute of Physics, The Hebrew University of Jerusalem, Jerusalem 91904, Israel 
\and NSF's National Optical-Infrared Astronomy Research Laboratory, 950 N. Cherry Ave., Tucson, AZ 85719, USA 
\and The University of Texas at Austin, 2515 Speedway Blvd Stop C1400, Austin, TX 78712, USA 
\and University of Massachusetts Amherst, 710 North Pleasant Street, Amherst, MA 01003-9305, USA 
\and Institute of Physics, Laboratory of Galaxy Evolution, Ecole Polytechnique F\'{e}d\'{e}rale de Lausanne (EPFL), Observatoire de Sauverny, 1290 Versoix, Switzerland 
\and Department of Physics and Astronomy, University of Louisville, Natural Science Building 102, 40292 KY, Louisville, USA 
\and Center for Astrophysics | Harvard \& Smithsonian, 60 Garden St., Cambridge MA, 02138, USA 
\and Black Hole Initiative at Harvard University, 20 Garden St., Cambridge MA, 02138, USA 
\and ESA/AURA, Space Telescope Science Institute, 3800 San Martin Drive, Baltimore, MD, 21218, USA 
\and Department of Physics and Astronomy, University of California, Los Angeles, 430 Portola Plaza, Los Angeles, 90095, CA, USA 
\and Aix Marseille Universit{\'e}, CNRS, CNES, LAM, 38, rue Fr\'ed\'eric Joliot-Curie 13388 Marseille CEDEX 13, France 
\and Kavli Institute for Cosmology, University of Cambridge, Madingley Road, Cambridge CB3 0HA, UK 
\and Cavendish Laboratory, University of Cambridge, 19 JJ Thomson Avenue, Cambridge CB3 0HE, UK 
\and Astronomy Centre, University of Sussex, Falmer, Brighton BN1 9QH, UK 
\and Institute of Space Sciences and Astronomy, University of Malta, Msida MSD 2080, Malta 
\and Departamento de Astronom\'ia, Universidad de La Serena, Av. Juan Cisternas 1200 Norte, La Serena, Chile 
\and ARAID Foundation. Centro de Estudios de F\'isica del Cosmos de Arag\'on (CEFCA), Unidad Asociada al CSIC, Plaza San Juan 1, E–44001 Teruel, Spain 
\and Laboratory for Multiwavelength Astrophysics, School of Physics and Astronomy, Rochester Institute of Technology, 84 Lomb Memorial Drive, Rochester, NY 14623, USA 
\and Department of Physics and Astronomy, Texas A\&M University, College Station, TX, 77843-4242, USA 
}
\date{Submitted to A\&A}

\abstract 
{
The galaxy integrated star-formation rate (SFR) surface density ($\Sigma_{\rm SFR}$) has been proposed as a valuable diagnostic of the mass accumulation in galaxies as being more tightly related to the physics of star-formation and stellar feedback than other star-formation indicators. In this paper, we assemble a statistical sample of $230$ galaxies observed with JWST in the GLASS and CEERS spectroscopic surveys to estimate Balmer line based dust attenuations and SFRs (i.e., from H$\alpha$, H$\beta$, and H$\gamma$), and UV rest-frame effective radii. We study the evolution of galaxy SFR and $\Sigma_{\rm SFR}$ in the first $1.5$ Billion years of our Universe, from redshift $z\sim4$ to $z\sim10$. We find that $\Sigma_{\rm SFR}$ is mildly increasing with redshift with a linear slope of $0.16 \pm 0.06$. We explore the dependence of SFR and $\Sigma_{\rm SFR}$ on stellar mass, showing that a star-forming 'Main-Sequence' and a $\Sigma_{\rm SFR}$ 'Main-Sequence' are in place out to $z=10$, with a similar slope compared to the same relations at lower redshifts, but with a higher normalization. We find that the specific SFR (sSFR) and $\Sigma_{\rm SFR}$ are correlated with the \xOIII/\xOII\ ratio and with indirect estimates of the escape fraction of Lyman continuum photons, hence they likely play an important role in the evolution of ionization conditions at higher redshifts and in the escape of ionizing radiation. We also search for spectral outflow signatures in the H$\alpha$ and \OIII\ emission lines in a subset of galaxies observed at high resolution (R$=2700$) by the GLASS survey, finding an outflow incidence of $2/11$ ($=20\%^{32\%}_{9\%}$) at $z<6$, but no evidence at $z>6$ ($0/6$, $<26\%$). Finally, we find a positive correlation between A$_V$ and $\Sigma_{\rm SFR}$, and a flat trend as a function of sSFR, indicating that there is no evidence of a drop of A$_V$ in extremely star-forming galaxies between $z \sim 4$ and $\sim 10$. This might be at odds with a dust-clearing outflow scenario, which might instead take place at redshifts $z\geq 10$, as suggested by some theoretical models.
}

\keywords{galaxies: evolution --- galaxies: high-redshift --- galaxies: ISM --- galaxies: star-formation --- galaxies: statistics}

\titlerunning{\footnotesize SFR and SFR surface density properties of galaxies from redshift $4$ to $10$}
\authorrunning{A.Calabr\`o et al.}
 \maketitle
\section{Introduction }\label{introduction}

The cosmic evolution of galaxy star formation rates (SFRs) is one of the fundamental predictions of astrophysical models and cosmological simulations, and one of the most studied processes observationally. Indeed, it provides essential insights into cosmic structure formation across all scales, the accretion of gas into these structures, the efficiency of conversion into stars, and ultimately the diffusion of baryonic material in the intergalactic medium (IGM) through stellar feedback \citep{white78,white91,springel03,shapley11,hopkins12,behroozi13,madau14}.

The SFR is intimately linked to other galaxy properties, the most important of which is stellar mass (\ensuremath{M_\star}). A correlation between SFR and \ensuremath{M_\star}, known as the `Main Sequence' of star-formation \citep{noeske07}, has been determined across over 5 orders of magnitudes in \mstar at all redshifts. While the slope $m$ of this relation remains rather constant with redshift, with $m$ ranging between $0.5$ and $1.2$, depending on both selection effects and the method to measure the SFRs (see \Citealt{speagle14} for a review), its normalization increases at earlier cosmic epochs. Indeed, focusing on the specific SFR (sSFR$=$SFR/\ensuremath{M_\star}), that is, the SFR normalized by the total stellar mass content, it follows a rather smooth, monotonic increase by at least one order of magnitude from redshift $0$ to the reionization epoch \citep{speagle14,pannella15,schreiber15,santini17,davidzon18,leslie20,popesso23}, which is supported by theoretical models and simulations \citep{furlong15,henriques15,behroozi19,donnari19,dicesare23}.

In addition to the SFR and the sSFR, a quantity that is gaining increasing attention now that JWST can resolve galaxy sizes out to the EoR, is the surface density of star-formation (\sigmaSFR). This quantity represents the SFR normalized by the surface area where it occurs, and encapsulates information about the spatial distribution of star formation (SF). It is usually defined by the expression \sigmaSFR $=$ SFR$/ (2 \pi \times r_e^2)$, where $r_e$ is the half-light radius. As galaxies become increasingly more compact at higher redshifts, since the dependence on the size is quadratic, \sigmaSFR increases more rapidly with redshift compared to the sSFR. 
In particular, it can rise by more than three orders of magnitudes in typical star-forming galaxies from $z\sim0$ to $z\sim6$ \citep{wuyts11,holwerda15}, reaching extreme conditions at the epoch of reionization (\sigmaSFR $\gtrsim 10$ \msun/yr/kpc$^2$).
Therefore, it is even more essential to characterize the peculiar conditions of SF in the first phases of galaxy assembly. 

Being more intimately related to the physics of star-formation, such as to the gas mass surface density through the Kennicutt-Schmidt relation \citep{kennicutt89,kennicutt07}, and to the effectiveness of stellar feedback, \sigmaSFR is thought to regulate the redshift evolution of the sSFR and \mstar \citep{lehnert15}. Moreover, for the same reason, \citet{salim23} introduced the term `\sigmaSFR Main Sequence' to indicate the \sigmaSFR vs \mstar relation, and they claim that this is even more fundamental than the sSFR - \mstar relation. Along this new `Main Sequence', star-forming galaxies have remarkably similar values of \sigmaSFR at $z\sim0$ (as also found independently by Schiminovich et al. 2007 and Forster-Schreiber et al. 2019), with only a weak dependence (i.e., a positive correlation) on stellar mass across over three orders of magnitudes from $10^8$ to $10^{11}$ \msun. This is indicative of a more universal and self-regulated nature of \ensuremath{\Sigma_{\rm SFR}}. This relation exists up to at least $z\sim2$ \citep{salim23}, even though it becomes slightly steeper compared to $z\sim0$, which they interpret as evidence of bulge formation in the central regions of more massive systems. 

The level of \sigmaSFR influences that of other galaxy properties. First, the increase in \sigmaSFR is tightly coupled to the increase of the interstellar medium (ISM) gas pressure and of the electron density $n_e$ with redshift \citep{jiang19,reddy23}. The increase of $n_e$ (and consequently \sigmaSFR) at fixed stellar mass may be entirely responsible for the higher ionization parameter of high redshift galaxies \citep{reddy23}. \sigmaSFR thus assumes the primary role that was previously attributed to metallicity. 
Secondly, \sigmaSFR is tightly related to the feedback of star-formation, regulating the power of the outflows and how efficiently the gas is removed from the galaxies, thus affecting the future evolution of the entire system. For instance, previous works have shown that \sigmaSFR correlates with the mass loading factor of the outflow \citep{llerena23}, and with the maximum outflow velocity  \citep{kornei12,bordoloi14,heckman16,sugahara19,prusinski21,calabro22a}. Similarly, an enhanced \sigmaSFR and stellar feedback reduce the gas and dust covering fraction in galaxies by creating channels that allow Ly$\alpha$ and Lyman-continuum (LyC) photons to escape into the IGM \citep{reddy22}.
This picture is supported by observations suggesting a relation between \sigmaSFR and the escape fraction \fesc of ionizing photons \citep{heckman01,naidu20,flury22}, according to which galaxies with higher \sigmaSFR at both low and intermediate redshift have larger fractions of LyC leakers, and also LyC leakers have higher \sigmaSFR than the average galaxy population. 

The key role of \sigmaSFR in driving outflows and easing the escape of ionizing radiation is also supported by models and simulations. 
For example, \citet{sharma16} and \citet{sharma17} implement galactic winds in the EAGLE cosmological simulations in such a way that they are capable of increasing \fesc to more than $20\%$ when \sigmaSFR reaches a value of $0.1$ \msun/yr/kpc$^2$. With this recipe, they are able to explain the redshift evolution of the average \fesc of galaxies and the volume filling factor of ionized gas up to $z\sim8$. 
\citet{naidu20} present an empirical model in which the evolution of \fesc in galaxies is only dependent on \ensuremath{\Sigma_{\rm SFR}}, which thus assumes a key role in reionization. With this tight connection between \fesc and \ensuremath{\Sigma_{\rm SFR}}, they claim that the bulk of reionization ($\sim80\%$) must be driven by a small number of galaxies ($<5\%$), the so-called ``oligarchs'', with extremely high \sigmaSFR ($\sim10$-$20$ \msun/yr/kpc$^2$), compact size, and relatively high UV luminosity (M$_\text{UV}<-18$) and stellar mass (\mstar $>10^8$ \msun).

The SFR in the first phases of galaxy assembly can also reach the super-Eddington regime, a condition in which the radiation pressure from young stars overcomes the gravitational force, leading to a galaxy-wide outflow that can clear the galaxy of its dust and gas content, pushing them away to large radii in a short timescale of the order of a few Myr \citep{ferrara23,ferrara24}. These extreme conditions of star-formation are invoked to explain the stunning abundance of UV luminous, very blue, and massive galaxies at $z>10$ found since the first extragalactic observations with JWST \citep[e.g.,][]{naidu22,castellano22}. 

The goal of this paper is to investigate how the SFR and the SFR surface density evolve in the first $1.5$ Billion years of our Universe. 
The investigation of galaxy assembly during the reionization epoch is perfectly-timed. On the one hand, the advent of JWST has allowed us to measure more compact galaxy sizes compared to previous telescopes, and to resolve internal structures at much finer levels. On the other hand, the unique sensitivity of JWST in the rest-frame optical and UV of galaxies has allowed us to perform large photometric and spectroscopic surveys, like GLASS \citep{treu22} and CEERS (Finkelstein et al., in prep.), targeting statistically representative samples of galaxies during reionization down to \mstar of $10^7$ \msun, and constraining the low SFR levels expected for these low-mass systems. This is essential to trace the scaling relations followed by the global galaxy population since the earliest epochs of their formation. 
In addition to the evolution of star-formation, we are interested in exploring how \sigmaSFR is related to other galaxy properties, and compare the observed values to those required by models and simulations to trigger galaxy wide outflows and to carve channels for the escape of Lyman continuum radiation. 

The paper is organized as follows. In Section \ref{methodology}, we describe the spectroscopic observations, sample selection, and the derivation of all the physical properties investigated in this paper. In Section \ref{sec:results}, we show our results, including the redshift evolution of the \sigmaSFR from $z=4$ to $z=10$, the SFR and the `\ensuremath{\Sigma_{\rm SFR}}' Main Sequence. Finally, in Section \ref{sec:discussion}, we compare our findings with theoretical model predictions, and explore the relation between \sigmaSFR and other galaxy properties to understand its role in the reionization process. Then we analyze the presence of outflows in NIRSpec galaxies and connect them to the star formation rate and dust attenuation. 
For our analysis, we adopt a Chabrier 2003 initial mass function (IMF), and we assume a standard cosmology with $H_{0}=70$ $\rm km\ s^{-1}Mpc^{-1}$, $\Omega_{\rm m} = 0.3$, $\Omega_\Lambda = 0.7$.

\section{Methodology }\label{methodology}

The aim of this work is to analyze fundamental scaling relations describing the evolution of star-formation in galaxies, hence it is important to assemble a statistical sample as large as possible. Here we use the data obtained by two JWST-based Early Release Science (ERS) programs, namely GLASS and CEERS, which so far represent the biggest publicly available JWST surveys, with hundreds of high-redshift galaxies observed in the optical and near-infrared rest-frame. In the following part we analyze each survey in more detail, and explain how we derive the physical properties of galaxies from the available spectroscopic, imaging, and photometric datasets. 

\subsection{Spectroscopic observations and sample selection}\label{sec:spectroscopic_observations_sample_selection} 

\begin{figure}[t!]
    \centering
    \includegraphics[angle=0,width=1\linewidth,trim={0.cm 0.cm 14cm 0.cm},clip]{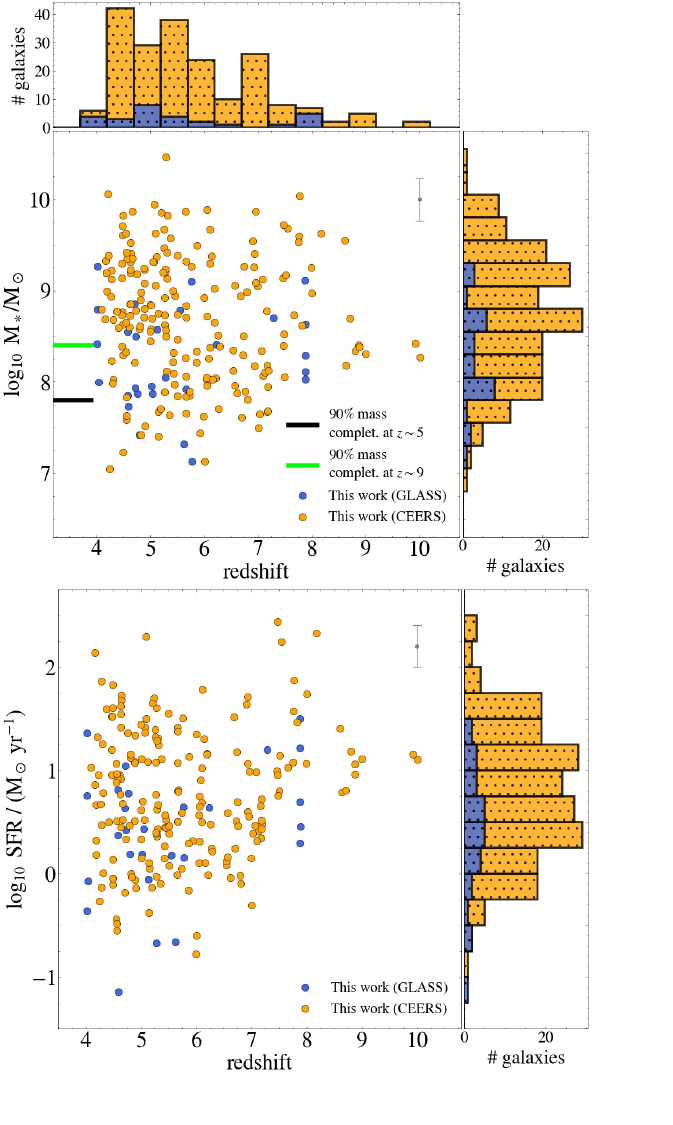}
    \vspace{-0.4cm}
    \caption{Distribution of stellar masses (top panel) and SFRs (bottom panel) for the galaxies selected in this paper. We highlight the redshift histogram of the entire sample on top of the first diagram, while the \mstar and SFRs histograms are added on the right part of each panel. Galaxies coming from the CEERS and GLASS surveys are differentiated with a yellow and blue color, respectively. We note that GLASS probes galaxies with slightly lower masses and SFRs on average, thanks to the gravitational lensing magnification effect. The gray vertical bars in the top right corners represent the typical uncertainties ($\sim0.2$ dex) for the stellar masses and SFRs. The $90\%$ mass completeness limits at $z\sim5$ and $z\sim9$ are highlighted, respectively, with black and lime horizontal segments.
    }\label{Fig:scatter_histograms}
\end{figure}

We first consider spectra acquired through NIRSpec MSA observations by the GLASS-JWST ERS Program (PID 1324, PI: Treu; \Citealt{treu22}) and by a JWST DDT program (PID 2756, PI: W. Chen; \Citealt{robertsborsani22}). Both of them targeted galaxies residing over the central regions of the Frontier Field galaxy cluster Abell 2744. However, they were obtained using two different setups and different exposure times. The former were taken in high-resolution grating mode with three spectral configurations (G140H/F100LP, G235H/F170LP, and G395H/F290LP), covering a wavelength range from $1.5$ to $5.14 \mu m$ at R $\sim 2000$-$3000$, and with a total integration time (per grating) of 4.9 hours. The latter instead were taken in low-resolution mode with the CLEAR filter + prism configuration, which provides a continuous wavelength coverage from $0.6$ to $5.3 \mu m$ at R $\sim 30$-$300$. The prism resolution in the red part of the spectrum is enough to resolve and detect redshifted rest-frame optical emission lines, such as \Hb\ and \xOIII. The exposure time in this case is of $1.23$ hours. The details of the spectral reduction, wavelength and flux calibration are described in detail in Mascia et al. (in prep.).  Since GLASS-JWST observations target sources behind a lensing cluster, we adopt the magnification factors ($\mu$) listed by \citet{mascia23}, based on the latest lensing model of \citet{bergamini23}, to correct line fluxes, SFRs, and galaxy sizes. 

We also consider the spectra obtained by the CEERS survey through the NIRSpec MSA mode. The CEERS spectroscopic data were acquired over $3$ different observing epochs and in $8$ different pointings. The NIRSpec pointings named p4, p5, p7, p8, p9, and p10 were observed with the three NIRSpec medium resolution (G140M, G235M, and G395M) gratings, covering the observed-frame wavelength range $0.97$-$5.10 \mu m$ with a resolving power R $\sim 1000$. In addition, p4, p5, p8, p9, p11, and p12 were also observed with the low resolution prism. In all cases, the exposure time is of $3107$s per spectrum. Three-shutter slitlets with a three-point nodding pattern were also used to enable accurate background subtraction. After the observing run, some galaxies have spectra obtained with both the prism and M-grating configuration, in which case we prioritize the latter, as the higher resolution of the M-grating improves the detection of faint emission lines with low S/N. The main steps of the reduction, the optimal extraction of the final 1D spectra, including also the wavelength and flux calibration, follow those described in other works of our collaboration, namely \citet{fujimoto23}, \citet{kocevski23}, and \citet{arrabalharo23}. A more detailed description of the NIRSpec data processing and a final spectroscopic catalog of the entire survey will also be presented in Arrabal Haro et al. (2024, in prep.). 

To build a sample of high redshift galaxies for our analysis, we visually checked all the 1D spectra provided by the above surveys in order to assign a first value to the spectroscopic redshift $z_{spec}$ based on the visual identification of relatively bright, optical emission lines, including \xOII, H$\beta$, \xOIII, and \Ha. This process was guided by the previous knowledge of the photometric redshifts of all the sources, obtained with the code {\sc eazy} \citep{brammer08}.
In the next section, we will describe a more precise measurement of $z_{spec}$. We also remark that the spectroscopic redshifts were checked and confirmed independently by other team members in both the GLASS and CEERS collaborations. Part of our spectroscopic sample is already published by other works \citep[e.g.,][]{arrabalharo23,mascia23,mascia24}, in which case we just checked the consistency of our $z_{spec}$ measurements. 

Among the sample observed with NIRSpec by either the GLASS or CEERS survey, we selected galaxies with a spectroscopic redshift $z_{spec}$ $\geq 4$. The reason of this lower limit is twofold. First, we aim at studying the properties of the galaxy population as we approach the epoch of reionization, where most of the scaling relations addressed in this work are still largely unexplored compared to Cosmic Noon ($1<z<3$). Secondly, targets with $z \geq 4$ were the main priority of the GLASS and CEERS spectroscopic surveys, thus we have better statistics in this range. 
This condition yields a final sample of $230$ galaxies that we consider in our analysis. Among this sample, $19$ have H-grating spectroscopic data (R$\sim2700$) in GLASS, $93$ have M-grating spectroscopy from CEERS (R$\sim1000$), while $15$ and $114$ are observed in prism configuration (R$\sim100$) in the Abell 2744 and EGS fields, respectively. The global population spans a redshift range between $z =4$ and $10.4$, with a median value $z_{med} = 5.4$ (see Fig. \ref{Fig:scatter_histograms}-\textit{top}). The number of sources decreases toward higher redshifts, with $11$ selected galaxies residing at redshift $\geq 8$. 

We can notice in Fig. \ref{Fig:scatter_histograms} that the GLASS sample includes $5$ galaxy members of a spectroscopically confirmed protocluster at redshift $\sim 8$ \citep{morishita23a}. Furthermore, $4$ galaxies in the same field have spectroscopic redshifts ranging $4.01$-$4.05$. All of them are more than $1 \arcmin$ away from each other, but the whole group lies within a projected radius of $80 \arcsec$ (i.e., $\sim 0.6$ pMpc at redshift $4$). This is much larger than the protocluster at $z \sim 8$, but more similar to the size of some galaxy overdensities identified at redshifts $\sim 3$ \citep[e.g.,][]{calabro22b}.

Given the variety of observational programs considered, it is useful to better investigate the properties of our selected sample to assess potential biases on our analysis. 
The GLASS observations with NIRSpec cover approximately $50$ arcmin$^2$ in the sky. Given the relatively long exposure times, they are among the deepest available so far, as we can detect emission lines with fluxes down to a few $10^{-18}$ erg/s/cm$^2$. In addition, the lensing magnification is ideal for improving the detection of the intrisically faint, low-mass galaxy population at the epoch of reionization. 
The primary science targets of the NIRSpec observations were placed into open Microshutters (MSA), and were chosen based on previous spectrophotometric catalogs available from the GLASS team and from the literature. Many of them already had a confirmed spectroscopic redshift, while the remaining systems were photometrically selected at $z>4$ using previous HST observations in the region of Abell 2744 (see \Citealt{treu22} for more details).
The CEERS spectroscopic observations cover instead a larger area in the sky of about $90$ arcmin$^2$, but they are shallower compared to GLASS. Some of the spectroscopic targets are HST-selected in the UV rest-frame (among which we find those galaxies falling outside of the NIRCam coverage), while additional targets benefitted from NIRCam observations undertaken in the first observing run (prior to NIRSpec observations), and therefore are optically selected, even though they might still be biased to UV-bright LBG-type objects as they are chosen based on their photometric redshift. While a more detailed description will be included in Finkelstein et al. (2024, in prep.), we note that it is difficult to precisely assess the selection function even within the same survey. It is thus a better approach to analyze a posteriori the galaxy properties as a whole. 

Even though GLASS probes slightly lower masses and SFRs on average compared to CEERS, due to the gravitational lensing effect, the two subsets are similarly distributed in redshift, as shown in the top histogram of Fig. \ref{Fig:scatter_histograms}-\textit{top}, and they are located along the same \mstar vs SFR relation. Moreover, in the redshift evolution of \ensuremath{\Sigma_{\rm SFR}}, and in all the other diagrams analyzed in this paper, removing the small GLASS subset does not significantly change the results and the best-fit relations, meaning that they satisfy the same physical trends. 
We also note from Fig. \ref{Fig:scatter_histograms} that both subsets may suffer from slight incompleteness at redshift $\geq 8$, where we may lose a fraction of galaxies at the lowest masses and SFRs (respectively $\log$ \mstar/\msun $<8$ and $\log$ SFR/(M$_\odot$/yr) $<0$. However, as will be shown later in Section \ref{sec:main_sequence_SFR}, our galaxies can be fairly considered as representative of the star-forming `Main Sequence' for most part of our redshift range. 

Finally, even though the standard STScI reduction pipeline adopted for our spectra already includes a slit loss correction, we further checked the absolute flux calibration of the output 1D spectra by comparing them to the available broadband photometry (see Section \ref{sec:stellar_masses} for the specific filters used). 
For all galaxies, we considered total fluxes estimated in the CEERS and GLASS collaborations. In the first survey, these fluxes are obtained by first performing the photometry in elliptical Kron apertures, and then applying aperture corrections estimated from simulations by matching the image PSFs between different filters, as detailed in \citet{finkelstein23}. A similar procedure for obtaining total fluxes was followed for the GLASS survey by \citet{paris23}, using the code \textsc{a-phot} \citep{merlin19}.
Despite the calibration corrections for individual photometric bands do not show a significant wavelength dependence, as also found by other works in the collaboration \citep[e.g.,][]{napolitano24}, we limit the comparison to the redder spectral range at $2 \mu m < \lambda < 5 \mu m$, as the S/N of the continuum rapidly drops at lower wavelengths, making this check and comparison more difficult. A second reason for this choice is that we are interested only in optical rest-frame lines, which mostly lie in the red part of the spectrum. 
For the correction, we apply to the spectrum a constant multiplicative factor, which is obtained as the median value of the corrections derived independently for all the photometric bands whose central wavelengths reside in the above spectral range.  

\subsection{Line measurements and derivation of star-formation rates}\label{sec:line_measurements}

We measure emission line properties with a similar methodology to that adopted in \citet{calabro19} and \citet{calabro23}. In brief, we use a python version of the $\chi^2$ minimization routine {\sc mpfit} \citep{markwardt09} to fit the following emission lines: \xOII, \Hg, \xOIII, \Hb, and \Ha, together with their underlying continuum in a spectral window of $\pm 5000$ km/s around the line. We assume a Gaussian function for the emission lines, while the continuum is modeled with a first order polynomial. We assume for the lines a common redshift and line velocity width $\sigma$, with a tolerance of $100$ ($1000$) km/s on the line centroid and $\sigma$ for the H- or M-grating and prism spectra, respectively. 
The fit yields the exact redshift of the sources, the line widths, the line fluxes, and their corresponding uncertainties. 
The quality of the fit is controlled through the reduced $\chi^2$ and visual inspection for each galaxy. We consider the lines as detected above a S/N of $3$, while, for non-detected lines, we adopt $3\sigma$ upper limits, which then are propagated into all the physical quantities derived. As discussed in \citet{calabro23}, the underlying stellar absorption for Balmer lines can be neglected as it does not produce significant variations of the emission line fluxes for galaxies in our stellar mass range. Moreover, our galaxies have typically younger ages, lower metallicity and lower dust content compared to the lower redshift CEERS sample analyzed in \citet{calabro23}, therefore the underlying absorption is expected to be even smaller \citep{groves12}. For this reason, we do not correct the emission line fluxes for stellar absorption. 
We also notice that in low resolution spectra acquired with the prism configuration, \Ha is blended with the [NII] doublet. However, our galaxies are all very low-mass and with sub-solar metallicities, for which the expected contamination is less than $10\%$, according to lower redshift studies \citep{faisst18}. A negligible contribution of \NII\ to \Ha for galaxies at $z>4$ is also found by the photoionization models described in \citet{simmonds23}, and supported by the results from our high resolution dataset. As a consequence, we do not attempt to make any correction on the \Ha flux in prism spectra. 
Finally, two sources at redshifts $5<z<6$ with clear broad \Ha\ components and narrow \xOIII, indicative of broad-line AGN emission, are removed from the sample. We refer to \citet{calabro23} for a more detailed discussion of the emission line measurement. 

As a first step of our analysis, we compute the gas-phase attenuation A$_V$ from the available Balmer lines (among \Hg, H$\beta$, and \Ha) assuming a dust-screen, Milky Way like extinction law \citep{cardelli89}. 
If only one Balmer line is detected, we adopt an A$_V$ as inferred from the SED fitting procedure described later in Section \ref{sec:stellar_masses}. 
As shown in the Appendix of \citet{calabro23}, hydrogen recombination lines across a large range of wavelengths ($ \sim 0.4$ to $\sim 2 \mu m$) are overall in agreement with the predictions of typical dust attenuation models, even when the lines reside in different gratings, indicating that their relative calibration and the inferred values of A$_V$ are generally reliable. 
From the fluxes of the \xOII\ and \xOIII\ lines, we also derive the O32 index, defined as the logarithm of the reddening corrected \xOIII/\xOII\ ratio.

We then derive the star formation rates as follows. We first compute the observed, dust-corrected \Ha luminosity L$_{H\alpha}$. If \Ha is not detected, we consider the fluxes of \Hb (or \Hg as last possibility), corrected for dust attenuation and rescaled to \Ha assuming the intrinsic ratios of $2.86$ and $6.10$ (respectively for \Hb and \Hg), valid for case B recombination, for a temperature $T=10000$K, and an electron density $n_e=10^2$ cm$^{-3}$ \citep{osterbrock89}. We then convert L$_{H\alpha}$ to SFR assuming the calibration of \citet{reddy22} as SFR $=$ L$_{H\alpha}$ $\times$ $10^{-41.67}$. This is the most suited calibration according to the typical subsolar metallicities expected for our galaxies at $z>4$, and reflects the greater efficiency of ionizing photon production in metal poor stellar populations. In our sample, we obtain the SFR from \Ha in $153$ galaxies, from \Hb in $67$ systems, while for $10$ galaxies it is inferred from \Hg. 
The resulting SFR distribution is shown in Fig. \ref{Fig:scatter_histograms}-\textit{bottom}: our galaxies span values between $\sim 0.1$ and $300$ \msun/yr, with a median SFR of $\sim 5$ \msun/yr. 

We note possible caveats of our analysis, affecting the relation between the SFR and \Ha luminosity. For example, the diffuse warm ionized gas (DIG) or collisional excitation in the ISM might contribute to the observed H$\alpha$ emission, leading us to slightly overestimate the SFRs. On the other hand, part of the ionizing radiation may be absorbed by dust and helium, or might escape from the galaxy, reducing the amount of Lyman continuum (LyC) photons available for hydrogen line emission. This might lead us to slightly underestimate the total SFR of the galaxy. Modeling and testing these effects is difficult, especially at high redshift. However, despite these uncertainties, the simple analysis performed in this paper and the dust correction through the Balmer decrement should work well as a first approximation, as shown by \citet{tacchella22}. 

We also note that the dust correction is another potential source of uncertainty. This is due both to a poor theoretical understanding of dust production mechanisms and dust growth in the ISM at high redshift (see discussion in \Citealt{markov23}), and to the lack of observational constraints. For this reason, different dust attenuation and extinction laws measured in the local Universe are usually adopted in the literature also at early epochs. We remark that our choice for dust correction is the same adopted by similar works on NIRSpec galaxies at the epoch of reionization \citep[e.g.,][]{llerena24}. Adopting a \citet{calzetti00} attenuation law, as done for example in \citet{mascia24}, does not significantly alter the results.

\subsection{Photometric data and estimation of stellar masses through SED fitting}\label{sec:stellar_masses}

We derive stellar masses for our galaxies through an SED fitting procedure using the code {\sc cigale} \citep{boquien19}, version 2022.1. We fit \citet{bruzual03} stellar population models with Chabrier IMF to the available observed photometry ranging from $0.5$ to $8 \mu m$. There is a homogeneous photometric coverage for all the galaxies in our sample. In particular, all the galaxies observed by the GLASS survey are covered by JWST/NIRCam imaging from the UNCOVER program (JWST-GO-2561; \Citealt{bezanson22}). For this subset, fluxes and uncertainties are obtained in the F115W, F150W, F200W, F277W, F356W, F410M, and F444W filters, using the catalog of \citet{paris23}. 
The majority of our NIRSpec sources in the EGS field are also covered by JWST/NIRCam observations from the CEERS survey in the same filters listed above \citep{bagley23}. For this subset, we take the fluxes and uncertainties from \citet{finkelstein23}. For all the NIRCam sources, we use total fluxes estimated as mentioned in Section \ref{sec:spectroscopic_observations_sample_selection}.
However, a subset of CEERS sources ($79$) do not have JWST imaging. 
To ensure also in this case a complete coverage of the UV and optical rest-frame emission for the SED fitting, we consider the following broadband filters: HST/ACS F606W and F814W, HST/WFC3 F125W, F140W, and F160W, CFHT/WIRCAM J, H, and K$_s$, Spitzer IRAC 3.6 and 4.5 $\mu m$. In particular, we use the total photometric fluxes and associated uncertainties from the multi-wavelength catalog 
assembled by \citet{stefanon17}. 

The SED fitting with {\sc cigale} is performed as follows. The redshift of the galaxy is fixed to the spectroscopic value estimated in Section \ref{sec:line_measurements}. We consider a main stellar population with metallicities Z$_\ast = 0.0004$, $0.004$, $0.008$, or $0.02$, and a delayed SFH parameterized by an e-folding time that can assume the following values: $10$, $100$, $500$, and $1000$ Myr, and with a grid of $18$ possible stellar ages ranging from $50$ Myr to $1.3$ Gyr. We also include the possibility of a recent exponential burst with SFH $\propto \exp ^{-t/\tau_{burst}}$, where $\tau$ can assume the values $1$, $20$, and $50$ Myr, while the ages can range from $5$ to $50$ Myr. We also allow the mass fraction of the late burst population to vary from $0$ to $80\%$ of the total mass formed in the entire galaxy history. We include contribution from nebular emission, where the nebular component is parametrized with an electron density $n_e$ of $100$ cm$^{-3}$, subsolar gas-phase metallicity equal to the stellar metallicity, and ionization parameter between $-3$ and $-2$. Finally, dust attenuation of the stellar and nebular components are modeled with a Milky Way extinction curve \citep{cardelli89}, and parametrized with a color excess E(B-V)$_{\rm stellar}$ ranging from $0$ to $3$ in steps of $0.05$, a total to selective extinction R$_V=3.1$, and a stellar continuum to nebular attenuation ratio of $0.44$ \citep{calzetti01}. Therefore, for each galaxy, an A$_V$ is computed as R$_V$ $\times$ E(B-V). The intergalactic medium (IGM) transmission is also taken into account following the model of \citet{meiksin06}. 
After the SED-fitting, we correct the stellar masses of the GLASS sources for the effects of lensing, using the magnification factors mentioned above. As demonstrated by \citet{furtak21}, this approach yields consistent results compared to correcting the photometry before the SED-fitting.

We find stellar masses \mstar ranging from $10^7$ to $10^{10.5}$ \msun (see Fig. \ref{Fig:scatter_histograms}-\textit{top}), with a nearly symmetric distribution around a median $\log_{10}$ \mstar/\msun $=8.7$ ($1 \sigma =0.7$). Combining these results with the SFR measurements described in the previous section, we find that the bulk of our galaxy population has sSFRs between $1$ and $100$ Gyr$^{-1}$, with median value of $10$ Gyr$^{-1}$. 
For galaxies fitted with E(B-V)$=0$, we also set an upper limit on their dust attenuation as E(B-V)$=0.05$, corresponding to our grid step and to the average uncertainties obtained with this methodology. 
Comparing the values of A$_{V,nebular}$ obtained from the SED fitting to those estimated from Balmer lines, we find that they are overall consistent with the $1:1$ relation, even though with a large scatter.

For galaxies with multiple coverage, we have also checked that removing NIRCam photometry and using only the remaining bands (HST + Spitzer + ground) in the SED fitting yields \mstar that are in agreement with estimates based on JWST photometry in the entire range spanned by our sample. Therefore, we do not introduce systematic biases (related to the measurement method) on the derived parameters for the subset that is not covered by JWST imaging. 
Our stellar masses are consistent with those obtained using the code {\sc zphot} \citep{fontana00} with a similar setup and SFH to our {\sc cigale} run, indicating that this quantity is rather robust and not significantly affected by the exact fitting procedure. 
Finally, we have also tested a nebular attenuation equal to the stellar attenuation, but this choice did not produce significant variations in the final results of this paper.

\subsection{Galaxy sizes and SFR surface density estimation}\label{sec:GALFIT_morphological_parameters_sigmasfr}

We measure the major-axis half-light radius $r_e$ of our galaxies by fitting their rest-frame UV images with the python software \textsc{galight} \citep{ding21}. 
A subset of $151$ galaxies is covered by JWST imaging, in which case we adopted the background-subtracted mosaics described in \citet{bagley23} and \citet{paris23}, for the CEERS and GLASS subset, respectively. For the remaining subset of $79$ galaxies (all in the CEERS field), we use instead HST imaging \citep{koekemoer11}. To homogeneously probe the same rest-frame UV window, we performed the measurements in the F115W band for galaxies at $z<5.5$, in the F150W filter for those at slightly higher redshifts ($5.5<z<7.5$), and in F200W for systems at $z>7.5$. In cases where only HST imaging is available, we adopted the F125W and F160W band for galaxies at redshifts lower and higher than $5.5$, respectively. 

To measure the galaxy sizes, we first produce $3\arcsec \times 3 \arcsec$ cutouts, which are given as input to the {\sc galight} code. All sources in the cutout, including the central galaxy of interest, are fitted simultaneously assuming that they are well represented by a Sersic profile. We constrain the following parameters to keep the fit within physically meaningful values: the Sersic index $n$ is free to vary from $0.2$ to $8.0$, and the axis ratio $q$ is set in the range $0.1<q<1$. After running {\sc galight}, sources are flagged according to the quality of the fit, with a flag $=0$ assigned to reliable fits, and flag $=1$ for sources that are not fitted well by the above procedure (indicative of a more complex shape), as checked visually in the observed vs best-fit luminosity profile that is output by the code. As a result, $213$ galaxies with a good size quality flag are considered when analyzing the \sigmaSFR in the following sections. 

The size uncertainties are assessed in accordance with \citet{yang22} and rescaled based on the signal-to-noise ratio derived from the photometry. Previous studies \citep[e.g.,][]{kawinwanichakij21} have shown that the outcomes from {\sc galight} are robust, and consistent with estimations made through conventional softwares like Galfit \citep{peng02}.
A subset of our galaxies are indistinguishable from point sources. In particular, we performed a subset of simulations, finding that $0.025 \arcsec$ and $0.1 \arcsec$ are the smallest measurable sizes in JWST and HST images, respectively. Therefore, we set these values as upper limits. 
In our sample, $64$ galaxies of $213$ ($\sim 30 \%$) are unresolved and have upper limits on $r_e$. We also notice that for the galaxies lying in the GLASS field (which are all covered by JWST imaging), 
the sizes are corrected for the effect of lensing. In particular, we divided the original sizes by $\sqrt{\mu}$ assuming an isotropic magnification, which represents a reasonable approximation considering the low magnification factors (median $\mu \simeq 2$) found in our sample.   

With the estimated sizes, we compute the galaxy-integrated SFR surface densities for all the objects. 
A subset of $25$ extremely compact galaxies have an upper limit on the size but also an upper limit on the SFR. We exclude these galaxies from the analysis as their \sigmaSFR would be totally unconstrained. As a result, the investigation of the redshift evolution of \sigmaSFR and its dependence on other galaxy properties will be made on a sample of $188$ galaxies. 
The distribution of $\Sigma_{\rm SFR}$ for this galaxy subset has a median value of $\log_{10} \Sigma_{\rm SFR}=0.6$ (in units of \mstar/yr/kpc$^2$, $1\sigma =0.8$), with individual systems spanning the logarithmic range between $-1.5$ and $2.5$.   

The UV rest-frame based $r_e$ values adopted in this paper probe the size of star-formation in the last $\sim100$ Myr (i.e., the continuum emission from young and massive stars), a slightly longer timescale than the \Ha (or \Hb) based SFRs, which are sensitive to the last $10$-$30$ Myr of star-formation. Ideally, more suitable emission line based sizes could be derived through IFU spectral observations or from imaging observations in photometric bands dominated by emission lines. To check whether the latter case occurs within our sample, we have estimated for all the galaxies the emission line contribution to the observed photometry. In particular, we have compared for all the available photometric bands the in-band fluxes obtained by integrating the observed spectrum across each filter transmission curve, and by summing the emission lines only. With this procedure, we find that the maximum emission line contribution to the photometry is of $\sim20 \%$ on average, exceeding $60\%$ in $9$ galaxies ($\sim 3.5 \%$ of the sample). However, for this small subset, the optical rest-frame sizes (in bands where the emission lines dominate) are in agreement with those estimated in the rest-frame UV, suggesting that using the latter does not significanly alter the final \sigmaSFR on average. Moreover, recent independent studies at similar redshifts based on JWST-NIRCam observations \citep[e.g.,][]{ono24,morishita23b} have shown that rest-frame UV sizes are in agreement with rest-frame optical sizes (where we might have contribution from strong optical emission lines).

\subsection{Estimating sample completeness and potential selection effects}\label{sec:completeness}

We derive in this section an approximate estimation of the mass completeness limits for our sample. The majority of our NIRSpec selected galaxies at redshifts $4 \leq z \leq 10$ is made of JWST-NIRCam selected sources within a magnitude limit in F277W of $\sim28.5$ mag. Using the recent results of \citet{cole24} and rescaling them to account for the different limiting magnitudes in F277W, we estimate a $90\%$ mass completeness limit of $10^{7.8}$ M$_\odot$ at $z\sim5$ and of $10^{8.4}$ M$_\odot$ at $z\sim9$ (i.e., $\sim0.2$ dex lower than the limits of \Citealt{cole24}). This redshift dependent completeness has effects on the stellar mass and SFR distributions shown in Fig. \ref{Fig:scatter_histograms}, and is likely responsible for the strong rise observed at $z > 7.5$ of the minimum stellar mass and SFR probed within our sample.

We also note that no additional selection effects are added to the spectroscopic sample by our analysis, as we have a SFR measurement (or at least an upper limit) and a size measurement from UV rest-frame imaging for all the galaxies selected in Section \ref{sec:spectroscopic_observations_sample_selection}. The galaxies without a measurement of \sigmaSFR (because of an upper limit on both the SFR and the size), or without a size (because of a complex shape) are randomly distributed in redshift and in the \mstar - SFR plane, so they do not produce systematic biases in the results.

We also investigate the effect of surface brightness dimming on source identification and sample selection. 
To this aim, we performed a set of Monte Carlo simulations, creating mock observations of galaxies in the F277W band following a similar approach to that adopted \citet{treu23} (see their Section 3.2). We placed the sources at different redshifts from $z=4$ to $z=10$ (in steps of $0.5$), assuming for simplicity a single Sersic light profile for the galaxies, with $n=1$ (consistent with observations of EoR galaxies, \Citealt{yang22}), $q= 1$, and $r_e$ ranging from $0.1$ to $1.5$ kpc (in steps of $0.05$ kpc). The background level was set to match the exposure time ($5207$ s) in CEERS, and then $50$ simulations were run for each configuration, applying each time a detection algorithm (using the photutils package in python) with a S/N detection threshold of $3$. 
From these simulations, we study how the detection rate varies as a function of redshift and $r_e$ for different total magnitudes of the source, from $28.5$ (the magnitude limit of our sample) to $27$, in steps of $0.3$ mag. 

We find that at $z\simeq9$, the maximum size for which a galaxy can be detected in F277W (at fixed total observed magnitude) is smaller than at $z\simeq4$ by $\sim0.2$-$0.25$ dex. However, galaxies become intrinsically smaller at higher redshifts. For example, Ward et al. (2023) have shown that the typical $r_e$ of galaxies decreases by a factor of $\sim 2$ ($0.3$ dex) from $z=4$ to $9$ (at fixed stellar mass), and also the mass - size relation contributes to have smaller sizes at high-$z$. This suggests that, within our `magnitude-limited' sample, size incompleteness does not play a significant role in the observed evolution of \sigmaSFR at higher redshifts. 
For example, assuming for galaxies at redshifts $8 < z < 10$ an average F277W magnitude of $\sim 28$, and a minimum detected SFR of $5$ \msun/yr (Fig. \ref{Fig:scatter_histograms}-bottom), we are able to probe \sigmaSFR values down to $\log_{10}$ \sigmaSFR/(\msun/yr/kpc$^2$) $\sim 0.4$, which is more than $2\sigma$ below the $z$ - \sigmaSFR relation extrapolated without the highest redshift bin.

\section{Results}\label{sec:results}

We explore in this section the redshift evolution of \sigmaSFR for our selected sample, and then focus on fundamental scaling relations involving \mstar, SFR, and \sigmaSFR in the entire redshift range between $\sim4$ and $\sim10$.
To characterize all the 2D distributions presented in this paper and quantify the level of correlation between two quantities, we adopt two different approaches throughout this work, both of which take into account (in a different way) the presence of upper and lower limits in the data. We explain them in detail in the next subsection for the redshift vs \sigmaSFR diagram, but the same steps of the analysis are followed for the other diagrams.

\subsection{Fitting methods}\label{sec:fitting_methods}

As a first approach to characterize the redshift evolution of \sigmaSFR, we compute the median redshift and the median \sigmaSFR of galaxies falling in different, well defined bins of increasing spectroscopic redshift ($4$-$5$, $5$-$6$, $6$-$7$, $7$-$8$, $>8$), which are shown as black empty diamonds in Fig. \ref{Fig:redshift_evolution}. In each bin, we also calculate the median absolute deviation (MAD) of \sigmaSFR, which is indicated with black error bars. In this calculation, we assign half of the \sigmaSFR value to galaxies with $3 \sigma$ upper limits, or twice the value in case of $3 \sigma$ lower limit on \sigmaSFR. 
In data science, imputation of upper (lower) limits with half (double) the detection limit is a common approach, and simulation studies have found that this is a better choice compared to assuming the detection limit itself or a zero value (for upper limits), as it introduces the least bias into the estimates \citep{beal01}.
Then, we derive a global, best-fit linear relation and $1 \sigma$ limits through Monte Carlo simulations. In detail, we perform $1000$ realizations of our data varying the data points according to their x and y axis $1 \sigma$ uncertainties. For each realization, we repeat the same process of median \sigmaSFR estimation in each of the previously defined redshift bins, and fit these median values with a first order polynomial. We finally derive the best-fit relation as the median slope and normalization of all the different realizations, including the uncertainties as their standard deviation.

In the second analysis method, we use the Python package PyMC \footnote{\url{https://peerj.com/articles/cs-1516/}; \url{https://doi.org/10.5281/zenodo.10371446}}. This is an advanced tool for statistical modeling featuring next-generation Markov chain Monte Carlo (MCMC) sampling algorithms such as the No-U-Turn Sampler (NUTS; \Citealt{hoffman14}), a self-tuning variant of Hamiltonian Monte Carlo (HMC; \Citealt{duane87}). HMC and NUTS can manage complex posterior distributions and fitting models, taking advantage of gradient information from the likelihood to achieve much faster convergence than traditional sampling methods.  
A first advantage of this approach is that it can be applied to the entire dataset without dividing the sample in multiple bins as in the previous case. A second advantage is that it implements the methods of the Bayesian survival analysis in the linear regression to deal with upper and lower limits on the y axis. 
Most importantly, this method does not impute the limits (i.e., does not strictly substitute them with a different value), but instead integrates them out through the likelihood. 
In pratice, the entire dataset is fitted with a linear relation, obtaining a first guess of the line parameters. Then, pyMC is run by assuming Gaussian priors for the slope and intercept, with $\sigma =1$ and the first guesses as centers. At this step, normal measurements, upper and lower limits are considered simultaneously in the fit.
Finally, the full posterior distributions of the slope and intercept are used to calculate the mean linear relation, the $1\sigma$ uncertainty, and for showing the ´posterior plot', that is, a stack of random draws from the posterior distribution of the slope and intercept. 
We report in Table \ref{table:results_correlations} the best-fit parameters obtained for the main relations studied in this paper and using both the linear regression methods described above. 

\subsection{The redshift evolution of the SFR surface density }\label{sec:redshift_evolution_sigmaSFR}

\begin{figure*}[t!]
    \centering
    \includegraphics[angle=0,width=0.97\linewidth,trim={0cm 0cm 0.cm 0cm},clip]{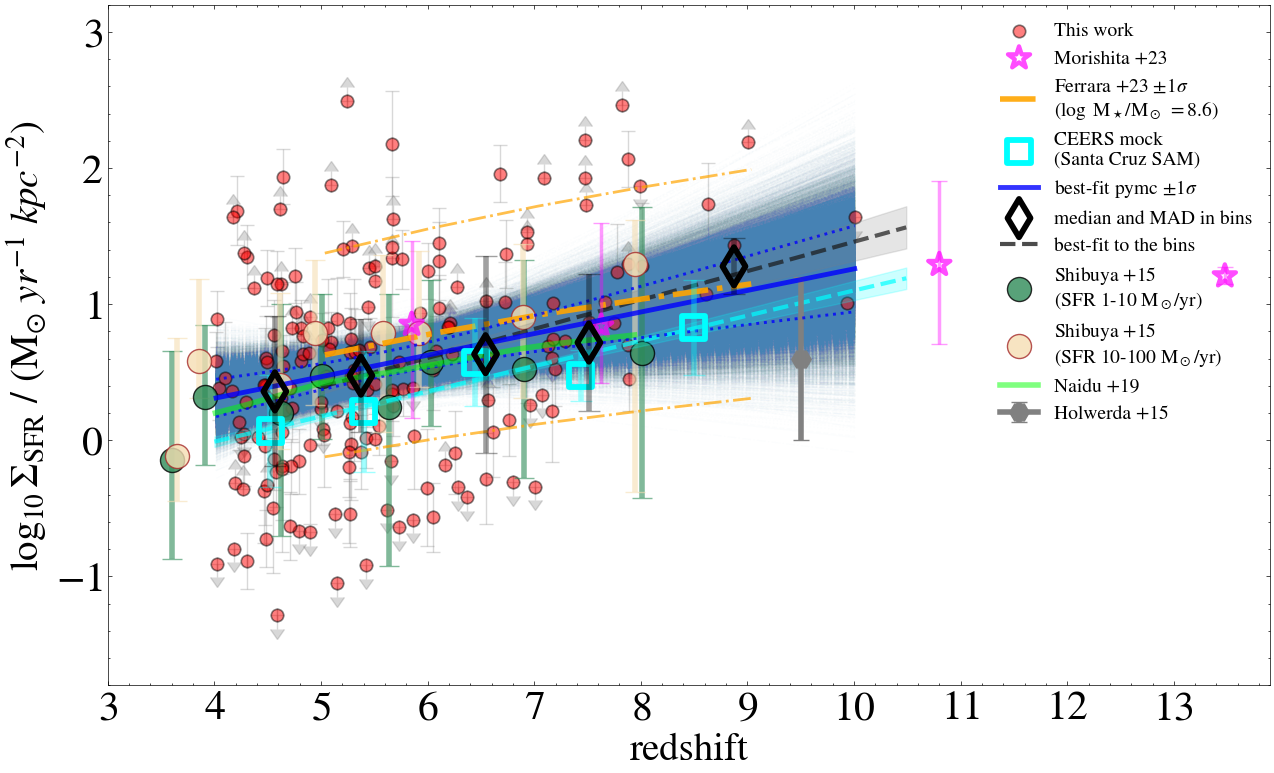}
    \vspace{-0.01cm}
    \caption{Redshift evolution of \sigmaSFR from $z=4$ to $10$. The median values and MAD of \sigmaSFR in bins of increasing redshift are shown with black empty diamonds and black vertical error bars, while the best-fit to the bins is represented with a black dashed line with a gray region indicating its $1\sigma$ uncertainty. The best-fit obtained with the Bayesian linear regression is shown with a blue continuous line, with the blue dotted lines and the blue shaded regions representing the $1\sigma$ uncertainty and the ´posterior plot', respectively.
    Theoretical model predictions and comparison observations are included as specified in the legend.   
    }\label{Fig:redshift_evolution}
\end{figure*}

\begin{table*}[t!]
\renewcommand{\arraystretch}{1.5} 
\vspace{-0.2cm}
\small
\begin{center} { 
\begin{tabular}{ | m{2.7cm} | m{2.95cm} | m{0.9cm} | m{2.5cm} | m{2.5cm} |} 
  \hline
  \textbf{x} & \textbf{y} & redshift range & best-fit bins  & best-fit pymc   \\ 
  \hline
  \hline
  $z$ & $\log$ $\Sigma_{SFR}$/(\msun/yr/kpc$^2$) & all &  $m=0.21 \pm 0.03$ \hspace{0.24cm} $q=-0.7 \pm 0.2$ & $m=0.16 \pm 0.06$ \hspace{0.24cm} $q=-0.3 \pm 0.4$ \\
  \hline
  $\log$ M$_\star$/\msun & $\log$ SFR/(\msun/yr) & low-z & $m=0.65 \pm 0.09$ \hspace{0.24cm} $q=- 4.9 \pm 0.8$ & $m=0.63 \pm 0.07$ \hspace{0.24cm} $q=-4.8 \pm 0.7$ \\
  \hline
  $\log$ M$_\star$/\msun & $\log$ SFR/(\msun/yr) & high-z & $m=0.4 \pm 0.3$ \hspace{0.34cm} $q=- 3 \pm 2$ & $m=0.63 \pm 0.10$ \hspace{0.24cm} $q=- 4.8 \pm 1.0$ \\
  \hline
  $\log$ M$_\star$/\msun & $\log$ $\Sigma_{SFR}$/(\msun/yr/kpc$^2$) & low-z & $m=0.17 \pm 0.13$ \hspace{0.24cm} $q=-1.0 \pm 0.9$ & $m=0.2 \pm 0.1$ \hspace{0.34cm} $q=-1.3 \pm 0.8$ \\
  \hline
  $\log$ M$_\star$/\msun & $\log$ $\Sigma_{SFR}$/(\msun/yr/kpc$^2$) & high-z & $m=0.2 \pm 0.4$ \hspace{0.36cm} $q=-1 \pm 3$ & $m=0.16 \pm 0.11$ \hspace{0.24cm} $q=-0.3 \pm 1.1$ \\
  \hline
\end{tabular} }
\end{center}

\vspace{-0.1cm}
\caption{\small Best-fit linear relations of the diagrams analyzed in this work. The first two columns indicate the $x$ and $y$ axis quantities, while the last two columns highlight the best fit coefficients of the linear relations ($y= m \times x + q$)  obtained with two different approaches, as described in the text. The third column indicates whether the relation is derived with the subset at lower redshift ($4<z<6$), at higher redshift ($6<z<10$), or in the entire $z$ range. Only galaxies above the $90\%$ mass completeness limits are used for the fit.
}
\vspace{-0.2cm}
\label{table:results_correlations}
\end{table*}

Our results for the evolution of \sigmaSFR are shown globally in Fig. \ref{Fig:redshift_evolution}, where the median \sigmaSFR and MAD in five increasing bins of redshift (defined as $4$-$5$, $5$-$6$, $6$-$7$, $7$-$8$, and $>8$) are represented with empty diamonds and vertical error bars, while the best-fit to the bins is highlighted with a black dashed line and a gray shaded area indicating its $1 \sigma$ uncertainty. In addition, we also consider the Bayesian fitting method, plotting the mean linear relation as a blue continuous line, the $1\sigma$ uncertainty as blue dotted lines, and the ´posterior plot' as a blue shaded region around the mean relation. 

Regardless of the analysis approach, we find a mild evolution of \sigmaSFR from $z \simeq 4$ to $z\sim 8$, with a slight but statistically significant increase by a factor of two ($0.3$ dex), going from $2.5$ \msun/yr/kpc$^2$ to $\sim 5$ \msun/yr/kpc$^2$. 
We note that in the short cosmic epoch that we are exploring ($\sim 1$ Billion years of cosmic history), the linear relation yields a good description of the increase in \sigmaSFR. In all cases we obtain a positive correlation 
with best-fit slopes of $0.16 \pm 0.06$ and $0.21 \pm 0.03$, respectively from the Bayesian regression and the fit to the bins. 
Therefore, in all cases we can exclude a flat trend at least at $2 \sigma$ level. 

Fitting the median \sigmaSFR in the five redshift bins, the slope is slightly higher compared to the other method, as due to a sudden rise of \sigmaSFR in the last bin at $8 < z < 10$, where we reach a median \sigmaSFR of $\sim 20$ \msun/yr/kpc$^2$. 
However, we notice that the median \sigmaSFR derived in the last bin is consistent with the Bayesian fit within $1 \sigma$. We have also removed the galaxies in the last redshift bin and fitted again the sample with the two methods, but we do not obtain significantly different results. Therefore the upturn of \sigmaSFR at $z>8$ is not significant from our data. 
This trend might be due to the different sample completeness limits between lower and higher redshifts, as discussed in Section \ref{sec:completeness}, and might reflect the rise observed at $z \gtrsim 7.5$ of the minimum stellar mass and SFR probed by our NIRSpec sample (Fig. \ref{Fig:scatter_histograms}). We also remark that in that redshift regime we are limited by the poor statistics, and a larger sample is required to better constrain the \sigmaSFR evolution at $z>8$.

Both the normalization and the slope of our \sigmaSFR - redshift relation are consistent with most of the previous observations. In particular, our Bayesian best-fit line falls exactly between the observed median relations estimated by \citet{shibuya15} up to redshift $8$ in two different bins of SFR ($1 <$ SFR/\msun/yr $<10$ and $10 <$ SFR/\msun/yr $<100$). Compared to these results, our findings indicate that the increase of $\Sigma_{SFR}$ can be extended up to at least redshift $10$. 
The median \sigmaSFR between redshift $4$ and $9$ are overall consistent with those calculated by \citet{morishita23b} with a sample similar to our study in terms of stellar masses and SFRs probed (fuchsia empty stars in Fig. \ref{Fig:redshift_evolution}). 
Moreover, NIRSpec galaxies ranging $8<z<10$ have \sigmaSFR comparable to those recently derived in photometrically selected systems observed by JADES at much higher redshifts ($10$-$13$) by \citet{robertson23}. 
Finally, our results are slightly higher, but still marginally consistent, than the range of \sigmaSFR ($1$-$20$ \msun/yr/kpc$^2$) measured by \citet{holwerda15} for $6$ galaxies at redshifts $9<z<10$ identified with HST in the CANDELS survey, even though these are intrinsically brighter systems compared to those analyzed in this work.

\subsection{The main sequence of star-formation at $z>4$ }\label{sec:main_sequence_SFR}

\begin{figure}[ht!]
    \centering
    \includegraphics[angle=0,width=0.99\linewidth,trim={0cm 1cm 14.7cm 0cm},clip]{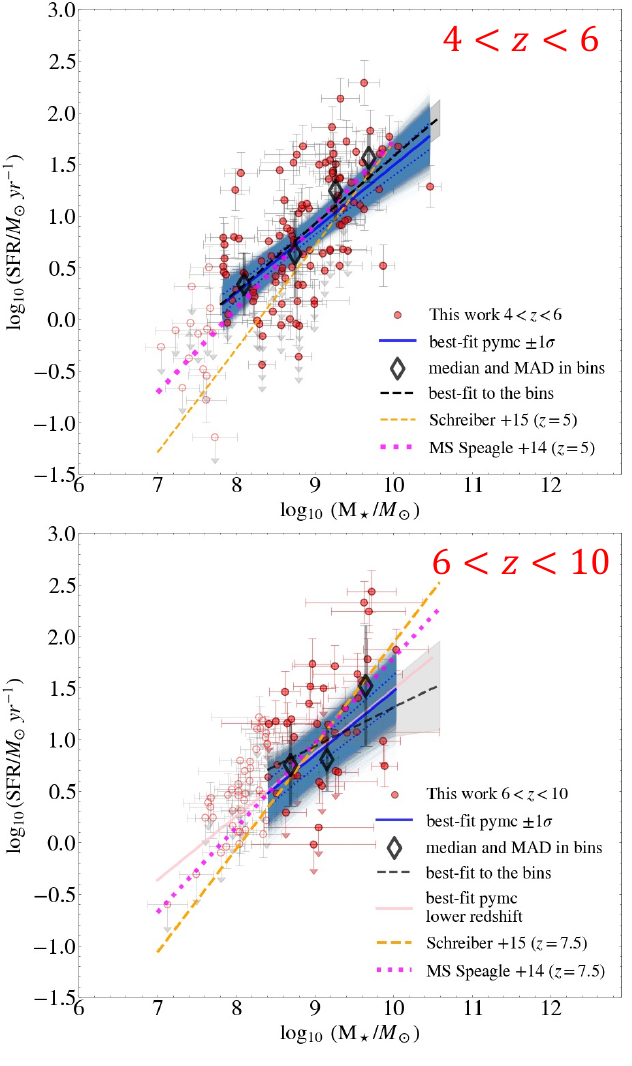}
    \vspace{-0.2cm}
    \caption{\textit{Top}: Diagram showing the star-forming `Main Sequence' of galaxies selected in this work in the redshift range $4 < z < 6$. The best-fit relations are derived in the mass range where we have $90 \%$ mass completeness. For comparison, the high redshift and low-mass extrapolation of the MS relations by \citet{schreiber15} and \citet{speagle14} (obtained at lower redshifts and higher \mstar) are shown with dashed orange and dotted fuchsia lines, respectively. \textit{Bottom}: Same as above, but for the redshift range $6<z<10$. The Bayesian best-fit relation obtained in the lower redshift bin in included as a pink continuous line for comparison.
    }\label{Fig:SFMS}
\end{figure}

We now focus on fundamental scaling relations describing the evolution of the galaxy population. The most important is that involving the star-formation rate and the stellar mass, which show a tight correlation at all redshifts, known as the star-forming `Main Sequence' (SFMS, \Citealt{noeske07}, \Citealt{daddi07}, \Citealt{santini17}). This relation is fundamental to understand the process of gas conversion into stars and subsequent build-up of stellar mass across cosmic time. While the slope is relatively constant with redshift, ranging from $\sim 0.6$ to $\sim1$, depending on the sample selection and specific tracer used for the SFR, all studies agree that the normalization increases monotonically from lower to higher redshift up to at least $z\sim 7$. 

In Fig. \ref{Fig:SFMS}, we show the \mstar vs SFR diagram for our selected spectroscopic sample. In order to check and better appreciate variations with redshift, we divide the sample in two redshift bins, that is, $4<z<6$ and $6<z<10$ (top and bottom panels, respectively). In both cases, we further divide the sample in different bins of stellar masses from $10^7.5$ to $10^{10.5}$ \msun in steps of $0.5$ dex (slightly larger in the first bin) when performing the first approach of the analysis explained in Section \ref{sec:fitting_methods}. We perform the fit considering only the galaxies above the mass completeness limits estimated in Section \ref{sec:completeness}, that is, $\log_{10}$ (\mstar/\msun) above $7.8$ in the low redshift bin, and $>8.4$ in the high redshift bin.
With all the fitting methods, we find in both redshift ranges a tight and significant correlation between \mstar and the SFR, indicating that the SFMS is in place up to the highest redshifts ($z\sim10$) explored by this work. Adopting the Bayesian regression as the reference method, the SFMS in the two redshift bins have similar slopes ($0.63 \pm 0.07$ at $z<6$ and $0.63 \pm 0.10$ at $z>6$), and the normalization of the relation is not significantly different between the two redshift bins in the \mstar range from $\sim 10^8$ to $\sim 10^{10.5}$ \msun.  

Recently, \citet{cole24} studied the SFMS using photometrically selected galaxies from the CEERS survey at redshifts $4.5$-$12$, with stellar masses and SFRs estimated through SED fitting. If we focus on the stellar mass ranges where we are both mass complete at $90 \%$ (i.e., $\log_{10}$ (\mstar/\msun) $>8$ at lower redshifts, and $>8.6$ at higher redshifts), we find consistent relations with similar slopes and normalizations. This also confirms that the spectroscopic selection is not significantly biased compared to a pure photometrical selection. 
We have also performed a fit by including all the galaxies down to the lowest measured stellar masses, but we obtain best-fit relations that are not significantly different from those estimated above. This suggests that, even though we are incomplete at lower stellar masses, this effect might not be large.

\citet{cole24} also found that 10Myr-averaged SFRs are significantly more scattered around the median relation compared to 100Myr-averaged values by $\sim 0.3$ dex (see their Fig.7), which they interpret as an evidence of the stochasticity of star-formation. We have here the possibility to analyze the scatter of the Main Sequence relation using the Balmer lines as tracers of very recent star-formation. To this aim, we computed the intrinsic scatter of the \mstar - SFR relations by subtracting in quadrature, for all the redshift and \mstar bins, the scatter due to the measurement uncertainty from the total observed scatter, as done in \citet{huang23}. We find that the intrinsic scatter $\log (\sigma_{\rm intrinsic, H\alpha}/(\msun/yr))$ ranges between $0.3$ and $0.6$ in both the low and high redshift bins, with a median of $0.5$ across the entire stellar mass range. This result is consistent with the range of $\sigma_{10 Myr}$ reported by \citet{cole24}, and $\sim0.4$ dex higher than $\sigma_{100 Myr}$ found by the same study, thus providing a spectroscopic confirmation of the relatively larger scatter of the Main Sequence when SFRs are averaged over short timescales of $\ll 100$ Myr.


Both the slope and normalization of our relations are generally in agreement with the extrapolations at redshift $5$ and $7.5$ (and down to a stellar mass of $10^7$ \msun) of the SFMS relations derived for more massive and lower redshift star-forming galaxies by \citet{speagle14} and \citet{schreiber15}, with a slightly better consistency with the shallower slope found by \citet{speagle14}. 
We note that the evolution of SFR at fixed stellar mass expected in this redshift range ($\sim0.15$ dex) is smaller than the intrinsic scatter of the relation and too small to be appreciated with our dataset, given the uncertainties obtained in the fit. 

\subsection{The main sequence of SFR surface density}\label{sec:main_sequence_sigmaSFR}

\begin{figure}[t!]
    \centering
    \includegraphics[angle=0,width=0.98\linewidth,trim={0cm 2cm 14.cm 0cm},clip]{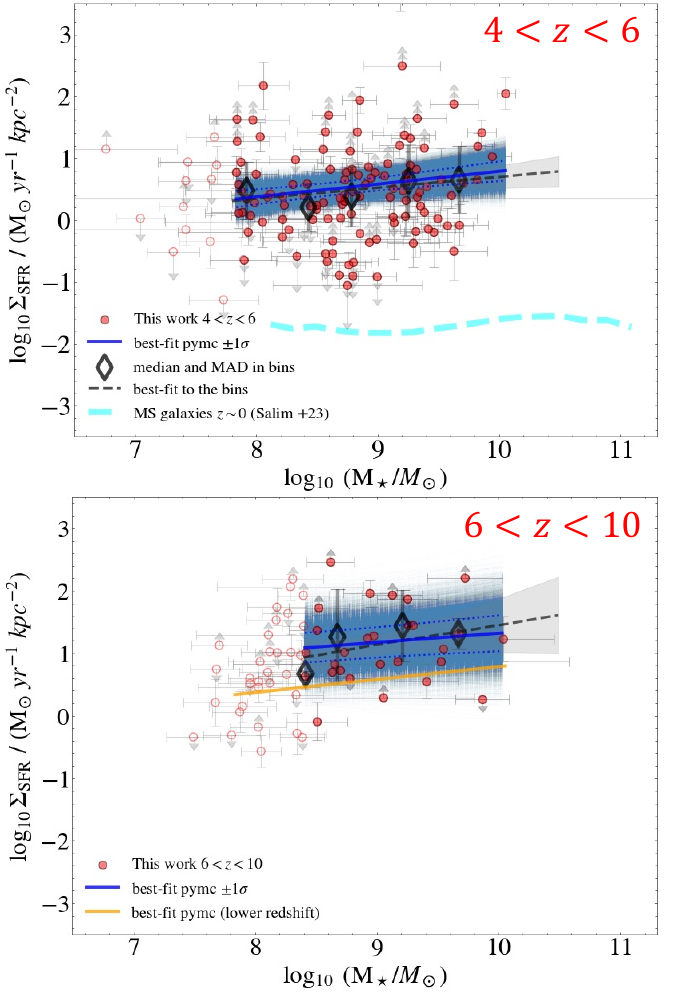}
    \vspace{-0.2cm}
    \caption{The \sigmaSFR\ `Main Sequence' for our selected galaxies in the lower redshift bin ($4<z<6$, \textit{top panel}) and higher redshift bin ($6<z<10$, \textit{bottom panel}). The lines, shaded areas, and markers are the same as in Fig. \ref{Fig:redshift_evolution}. The average location of massive (\mstar $> 10^9$ \msun) star-forming galaxies at redshift $\sim 0$ presented by \citet{salim23} is highlighted with a cyan dashed line. 
    }\label{Fig:sigmaSFR_MS}
\end{figure}

The galaxy integrated \sigmaSFR has been proposed as an alternative and more physical measurement of the current SF activity of galaxies, the reason being a closer connection to the gas density and the efficiency of stellar feedback compared to the specific SFR. For example, the incidence of outflows has been shown to correlate better with the \sigmaSFR\ rather than with the offset from the standard SFMS \citep{forsterschreiber19}. Moreover, \sigmaSFR\ also incorporates, by construction, information about the galaxy structure and morphology. 
For this reason, an alternative `Main Sequence' relation has been defined as the main locus occupied by star-forming galaxies in the \sigmaSFR - \mstar plane. Compared to the sSFR, the \sigmaSFR\ properties of galaxies are more uniform at varying stellar masses, such that \sigmaSFR\ is approximately constant across the Main Sequence, as found by \citet{schiminovich07} and \citet{salim23} for relatively low redshift galaxies ($z \lesssim 1$). 
However, there could also be deviations from this flat relation, indicating additional (or lower) star-formation activity associated with different galaxy structures. For example, moderately positive correlations between \sigmaSFR and \mstar are observed at \mstar $>10^9$ \msun by \citet{salim23} for star-forming galaxies from $z=0$ to $z=2$, which are interpreted as the build-up of a central bulge component in those galaxies.

We now explore the \sigmaSFR\ vs \mstar\ diagram for the NIRSpec galaxies, divided in two redshift bins ($4<z<6$ and $6<z<10$), as done for the classical SFMS. In the fit we consider only the galaxies above the mass completeness limits defined before for each redshift. Our results are shown in Fig. \ref{Fig:sigmaSFR_MS}. 
We find that in both redshift bins there is a very mild correlation, with \sigmaSFR slightly increasing in more massive galaxies: the best-fit slopes obtained with the Bayesian regression method are of $0.19 \pm 0.09$ and $0.30 \pm 0.15$, respectively in the low and high redshift bins. 
This tells us that the data points are still consistent (within $3 \sigma$) with a constant \sigmaSFR as a function of \mstar in the entire redshift of our work. Considering the large range spanned by \sigmaSFR (by more than three orders of magnitude), it is indeed remarkable to observe a median variation of only less than $0.5$ dex across more than $\sim 3$ orders of magnitude of stellar mass, from $\log$ \mstar $=7.5$ to $\sim10.5$.  For the higher redshift sample we also find a slightly higher normalization of the \sigmaSFR-`Main Sequence' by $\sim0.5$ dex (at \mstar $=10^9$ \msun) compared to the subset at $z<6$. 

The very mild increase of \sigmaSFR with \mstar is in agreement with previous observations of star-forming galaxies at comparable masses but at lower redshifts \citep{forsterschreiber19,schiminovich07,salim23}. In such comparisons, we do not find clear evidence of a sudden or significant upturn of the relation in the entire mass range that we explore. 
However, the normalization of the \sigmaSFR-`Main Sequence' is significantly higher compared to lower redshift studies. The median \sigmaSFR of our galaxies at $z \sim 5$ is of $\sim 2.5$ \msun/yr/kpc$^2$ ($\log$ \sigmaSFR $=0.4$). At fixed stellar mass, and focusing on the overlapping stellar mass region ($8 <  \log$ \mstar/\msun $<10$), our \sigmaSFR values are higher compared to the local Universe by approximately $2$ dex at $z \sim 5$ and $2.5$ dex at $z\sim 10$, indicating a strong evolution of \sigmaSFR with cosmic time.  
This is consistent with the average \sigmaSFR evolution found over this cosmic epoch by other observations \citep[e.g.,][]{shibuya15} and simulations \citep[e.g.,][]{sharma17}. It can be explained as the combined effect of the higher average SFRs (by $\sim1$ dex, \Citealt{speagle14}) and more compact sizes ($r_e$ $\sim 0.5$ dex lower, \Citealt{ward24}) of galaxies at $z\sim5$ compared to the local Universe, at a fixed stellar mass of $\sim10^{9.5}$ \msun.

\section{Discussion}\label{sec:discussion}

In this section we further analyze our results to better understand how the fundamental quantity \sigmaSFR is related to other phsyical properties of the galaxies during and immediately after the epoch of reionization.

\subsection{Comparison to predictions of theoretical models}\label{theoretical_models}

We compare our observations to some theoretical models that have been introduced to explain galaxy evolution across cosmic epochs. 
Focusing on the redshift evolution of \sigmaSFR, we first consider the empirical model of \citet{naidu20}. As already mentioned in Section \ref{introduction}, the main assumption of this model is that \fesc is solely dependent on \sigmaSFR as \fesc $=$ a $\times$ $\Sigma^b_{\rm SFR}$, with $a=1.6\pm0.3$ and $b=0.4\pm0.1$.  
As shown in Fig. \ref{Fig:redshift_evolution}, this model yields an evolution of \sigmaSFR up to redshift $\sim 8$ (lime curve) that is fully in agreement with our best-fit relation. 

As an additional test, we compare our observations to the physical model presented by \citet{ferrara23} and \citet{ferrara24}. In detail, we take the redshift evolution of the SFR predicted by their model, and derive the \sigmaSFR by combining it with the empirical size evolution found by \citet{morishita23b}. 
At the median stellar mass of the sample plotted in Fig. \ref{Fig:redshift_evolution} ($\log$ \mstar/\msun $=8.6$), we obtain a slightly increasing trend of \sigmaSFR, highlighted with a dash-dotted orange line, whose slope is remarkably consistent with our data and with the model of \citet{naidu20}. Even though is has a slightly higher normalization of $\sim +0.2$ dex, this is well below the uncertainties of our best-fit relation. 
We note that the prediction is very sensitive not only to the stellar mass considered (lower \mstar have systematically lower \sigmaSFR), but especially to the effective radius. Indeed, considering the uncertainties on the best-fit redshift-size relation of \citet{morishita23b}, we would obtain an uncertainty on the normalization of the predicted \sigmaSFR evolution of $\sim 0.8$ dex, higher than the median absolute deviation (MAD) of our sample.

Finally, we also take the CEERS mock galaxy catalog \citep{yung22} for comparison. This catalog covers an area overlapping with the observed EGS field, and contains galaxies over $0 < z \lesssim 10$. The galaxies in the mock lightcone are simulated with the Santa Cruz SAM for galaxy formation \citep{somerville15,somerville21}. The free parameters in the model are calibrated to reproduce a set of galaxy properties observed at $z\sim 0$ and have been shown to well-reproduce the observed evolution in high-redshift ($z \gtrsim 4$) one-point distribution functions of $M_{\rm UV}$, $M_\ast$, and SFR \citep{yung19a,yung19b}. The effective radii of these simulated galaxies are determined as described in \citet{brennan15} and are shown to be in good agreement with CEERS observations at $3 < z < 6$ \citep{kartaltepe23}.

From the mock lightcone, we extract a random sample of galaxies that has a redshift and a stellar mass distribution similar to our observed sample, and then apply a magnitude cut as F277W $<29.15$ to match the depth of the CEERS survey \citep{finkelstein23}, where most of our galaxies come from. With this selection method, fitting the median \sigmaSFR in five bins of redshift as done for the real data, we obtain an increasing trend of \sigmaSFR with redshift that is very similar to the observational result. In addition, the SAM best-fit relation has a slightly lower normalization than the observed one by $\sim0.2$ dex. As in the previous case, this is due to predictions on galaxy sizes, which are slightly larger on average compared to our values and to the size evolution obtained by \citet{morishita23b}. Despite this, the difference in \sigmaSFR normalization is rather small and below the $1\sigma$ uncertainties. Therefore, the SCSAM yields a good representation of the observed properties of galaxies in the entire redshift range from $4$ to $\sim 10$.
A similar good consistency between SCSAM model predictions and observations is obtained also for all the other diagrams studied in Section \ref{sec:results} and in Section \ref{sec:discussion}. 
The models analyzed in this section also agree on the strong evolution of \sigmaSFR from $z=0$ to the reionization epoch mentioned in Section \ref{sec:main_sequence_sigmaSFR}.
To conclude, our results are consistent not only with previous observations, but also with a variety of theoretical models that are commonly adopted in the literature.

\subsection{SFR surface density and ionization properties}\label{discussion:ionization}

\begin{figure}[t!]
    \centering
    \includegraphics[angle=0,width=0.98\linewidth,trim={0cm 0cm 0cm 0cm},clip]{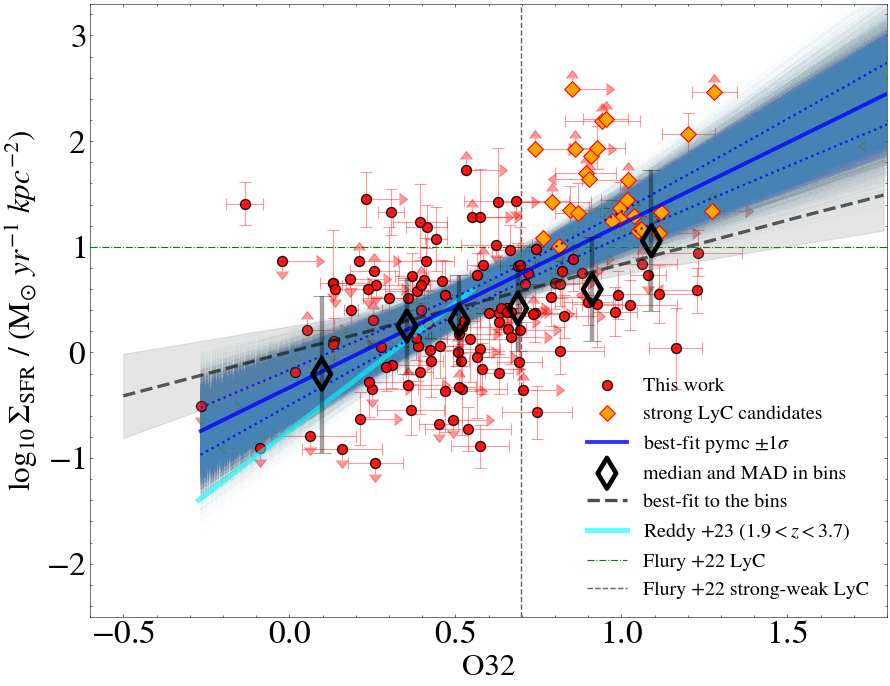}
    \vspace{-0.2cm}
    \caption{ \sigmaSFR as a function of the O32 index for our selected sample in the entire range $4<z<10$. The lines, shaded areas, and markers are the same as in Fig. \ref{Fig:redshift_evolution}. The diamonds colored in orange highlight our strong LyC leaking candidates, according to the criteria defined by \citet{flury22}.
    }\label{Fig:sigmaSFR_O32}
\end{figure}

The SFR surface densities of galaxies have been shown recently to be tightly related and to have a key influence on their ionizing properties \citep{reddy22}.  
We know from simulations, galaxy models, and observations, that high redshift galaxies have higher ionization parameters (at fixed stellar mass) compared to local analogs \citep[e.g.,][]{brinchmann08,nakajima13,steidel14,kewley15,shapley15,hirschmann17,hirschmann23,scharre24}.
This redshift evolution of ionization conditions is usually explained as a combined effect of lower metallicity, harder stellar ionizing spectra, and higher gas densities at $z>>0$. 
The latter increasing trend, also found with the electron density $n_e$ \citep{isobe23}, which is more easily measurable from optical emission lines, reflects the denser, more compact molecular clouds, and ultimately also higher values of \sigmaSFR. Significant correlations were indeed found observationally between $n_e$ and \sigmaSFR \citep{shimakawa15,jiang19,reddy23}, confirming the close relation between the two quantities, and the important role of \sigmaSFR in the evolution of the ISM properties across cosmic time (see also \Citealt{papovich22}). 

We test here the connection between \sigmaSFR and the ionization conditions, comparing the measured values of \sigmaSFR to the O32 index defined in Section \ref{sec:line_measurements}, which is usually adopted as an observational proxy for the ionization parameter. 
The result is shown in Fig. \ref{Fig:sigmaSFR_O32}. The O32 values, represented on the y-axis, range between $-0.3$ and $1.3$, with a median of $\sim 0.6$. We observe an increase of O32 with \sigmaSFR, with a slope of $1.5 \pm 0.2$, indicating that the correlation is statistically significant at more than $3 \sigma$ level. We notice also that the galaxies in the top right part of the diagram, with \sigmaSFR $\geq 1$ and O32 $\geq 0.7$, have similarly high \sigmaSFR and O32 to photometrically selected extreme emission line galaxies (EELGs) identified in the CEERS field by \citet{llerena24}. We refer to that paper for a more detailed discussion of those galaxies. 
Overall, these findings corroborate previous results, supporting the close relation between \sigmaSFR and the ionization properties of galaxies. We also note that excluding the small subset of galaxies at $z>8$ with very high \sigmaSFR does not significantly change this global picture.

Finally, $7$ galaxies observed at high resolution in GLASS have a S/N of \xOII $>3$ and a physically meaningful flux ratio \OII$_{3729}$/\OII$_{3726}$ in the range $0.3$-$1.5$ \citep{osterbrock89}. From this flux ratio, we derived for those galaxies the electron density $n_e$, using the python package \textsc{pyneb} \citep{luridiana15} and assuming an electron temperature $T=10000$K, as above. We find that most of the galaxies cluster within a range of densities going from $200$ to $400$ cm$^{-2}$ (median $n_e$ $=280$ cm$^{-2}$). The poor statistics does not allow us to test possible correlations between $n_e$ and \sigmaSFR, even though we notice that the galaxy ID 100003 with the highest density in the sample ($> 10^4 cm^{-2}$, the only one outside the range $200$-$400$ cm$^{-2}$), also has the highest \sigmaSFR ($\sim 10^2$ \msun/yr/kpc$^2$). We defer a more detailed analysis to a future work with a larger sample.

\subsection{SFR surface density and escape of Lyman continuum photons}\label{discussion:SigmaSFR_fesc}

\begin{figure}[ht!]
    \centering
    \includegraphics[angle=0,width=0.97\linewidth,trim={0cm 0cm 0.cm 0.0cm},clip]{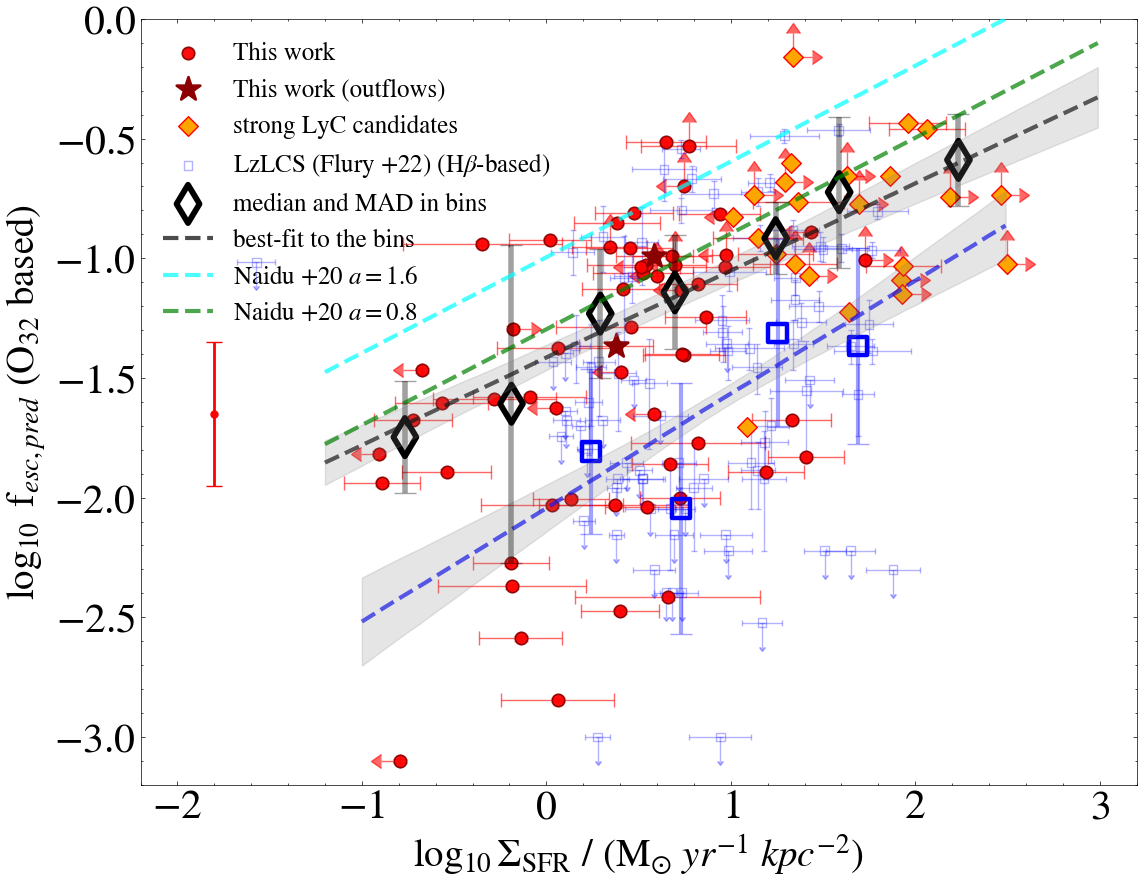}
    \vspace{+0.1cm}
    \includegraphics[angle=0,width=0.98\linewidth,trim={0cm 0cm 0.05cm 0.0cm},clip]{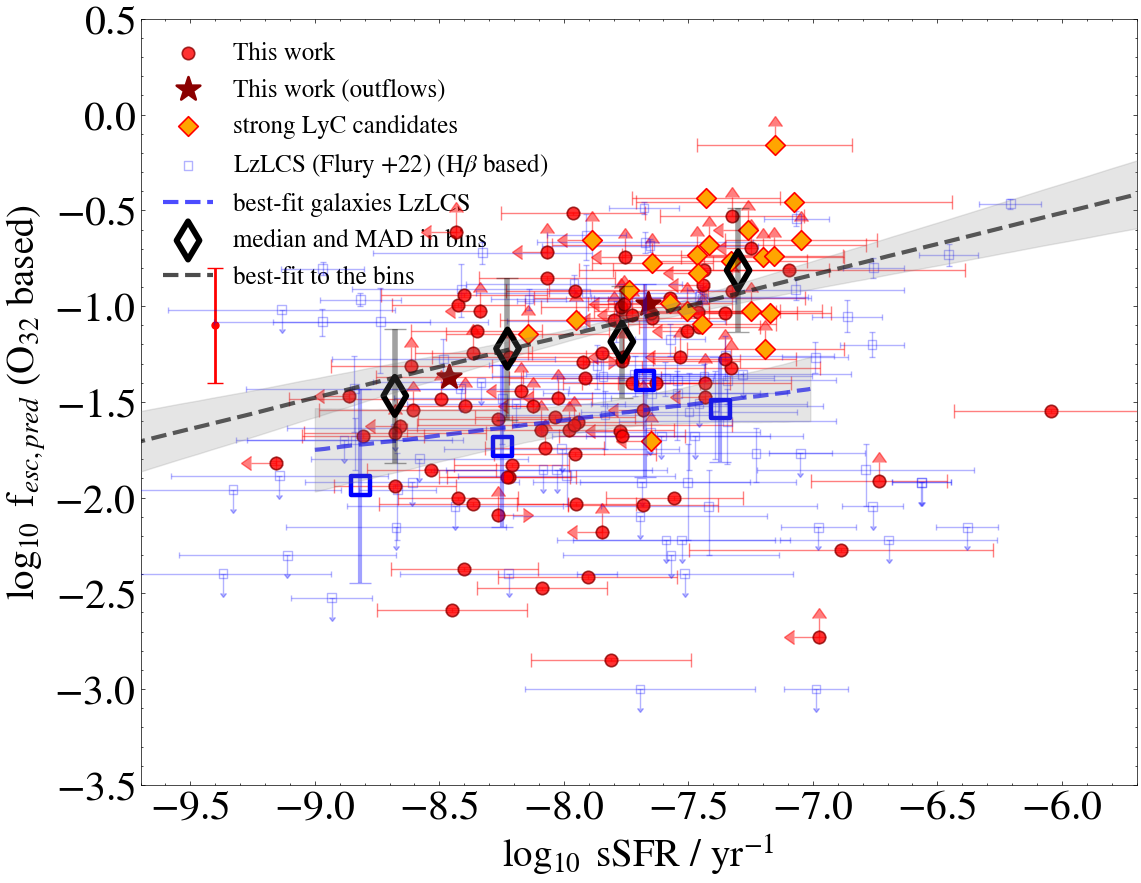} 
    \vspace{-0.05cm}
    \caption{ \textit{Top:} \sigmaSFR vs \fesc for galaxies at $4<z<10$ with an estimate of \fesc. The lines, shaded areas, and markers are the same as in Fig. \ref{Fig:redshift_evolution}. \textit{Bottom:} sSFR vs \fesc for this NIRSpec sample. The orange diamonds represent strong LyC leaking candidates according to \citet{flury22}, as highlighted in Fig. \ref{Fig:sigmaSFR_O32}. The two galaxies for which we detect outflow signatures in their H-grating spectra from GLASS are shown as big darkred stars. The sample representative error of $0.3$ dex on \fesc is added below the legend. We include with blue empty squares the LzLCS sample of $66$ galaxies at $0.2<z<0.4$, with measurements (including direct estimates of \fesc) performed by \citet{flury22}. 
    }\label{Fig:sigmaSFR_fesc}
\end{figure}

Enhanced values of the O32 index and \sigmaSFR are usually associated with increased fractions of ionizing photons that escape from the galaxies. 
The connection between \sigmaSFR (or O32) and \fesc is found from multiple observations and with different approaches. 
Lyman continuum leaking sources (LyC), identified through direct or indirect methods, are found to have \sigmaSFR\ significantly higher than the average value of the star-forming population at a given redshift \citep[e.g.,][]{cen20}. In the other direction, galaxies with enhanced \sigmaSFR\ show larger fractions of LyC leakers. For example, \citet{heckman01} and \citet{heckman02} claim that galaxies with \sigmaSFR above $0.1$ \msun/yr/kpc$^2$ are capable to launch strong star-forming driven winds, which might be responsible for creating channels though which Lyc photons can leak.
\citet{steidel18}, based on a sample of LBGs at $z\sim3$, suggest that \sigmaSFR and \fesc are highly correlated quantities.
Similarly, the O32 index is considered as a valid indicator of LyC leaking, according to \citet{izotov20}. \citet{flury22} also found that $40\%$ of local galaxies with \sigmaSFR$>10$ \msun/yr/kpc$^2$ and O32 higher than $0.7$ are LyC leakers, hence they adopted these two conditions at higher redshifts to define a source as a strong LyC leaker candidate. 

Following these observational findings, hydrodynamical simulations and models have started to implement the effects of spatially concentrated star-formation in the form of turbulence and stellar-driven winds \citep{ma16,sharma16}, showing that an increased \sigmaSFR is responsible for launching galaxy-wide outflows and for carving ionizing channels in the ISM where the Lyman continuum radiation of young massive stars and supernovae is free to propagate outside. 
According to the model of \citet{naidu20}, \fesc reaches unity when \sigmaSFR is one-third of the maximum \sigmaSFR ($=1000$ \msun/yr/kpc$^2$) that can be sustained without radiation pressure instabilities \citep{hopkins10}.

Triggered by this extensive observational and theoretical background, we explore with our dataset the relation between \sigmaSFR and \fesc at redshifts $4<z<10$. At $z\geq4$ it is almost impossible to directly detect the Lyman continuum emission due to the very low IGM transmission \citep{vanzella15}, and at $z>5$ it becomes virtually impossible.
For this reason, we determine \fesc in an indirect way using the empirical relation presented in \citet{mascia23}, which provides an estimation of the escape fraction using three galaxy parameters: the UV beta slope, the physical size $r_e$, and the O32 index, which are amongst the properties better correlated with \fesc. 
This relation is calibrated on the low-redshift ($0.2 < z < 0.4$) Lyman Continuum survey (LzLCS), for which \citet{flury22} performed direct measurements of the Lyman continuum flux and \fesc from imaging observations.

In Fig. \ref{Fig:sigmaSFR_fesc}-\textit{top}, we show \sigmaSFR as a function of \fesc. Using our two analysis approaches, we find in all cases a significant correlation between these two parameters, according to which galaxies with higher \sigmaSFR have systematically higher \fesc than galaxies with lower \sigmaSFR. In detail, \fesc increases from $\log$ \sigmaSFR $=-1$ to $3$, with a best-fit slope of $0.37 \pm 0.05$ by fitting the median values in $7$ bins of \sigmaSFR defined with the following grid [$-1$, $-0.5$, $0$, $0.5$, $1$, $1.5$, $2$, $3$]. These results do not significantly change if we exclude from the analysis the galaxies at $z > 8$ that are responsible for the \sigmaSFR enhancement in the last redshift bin observed in Fig. \ref{Fig:redshift_evolution}.
This indicates that galaxies with more concentrated SFR have the conditions that should facilitate the escape of LyC photons. 
A subset of $\sim 20\%$ of our galaxy sample have \sigmaSFR and O32 satisfying the conditions established by \citet{flury22} for being strong LyC leaker candidates, showing similar properties to the LyC leakers studied at redshift $\sim 0.3$. These systems, indicated with orange diamonds in Fig. \ref{Fig:sigmaSFR_fesc} and in Fig. \ref{Fig:sigmaSFR_O32}, tend to have higher \fesc estimates compared to the average galaxy population at these redshifts. 

We compare our observations to the direct measurements of \fesc and physical properties estimated by \citet{flury22} for the LzLCS galaxy sample (blue empty squares in Fig. \ref{Fig:sigmaSFR_fesc}). We find a similar slope of the best-fit relation, even though we derive a higher normalization by $\sim 0.5$ dex. 
The large difference in normalization of the \sigmaSFR - \fesc relation mostly depends on the fact that, at fixed \sigmaSFR, high redshift galaxies have a lower UV slope on average compared to the low redshift systems analyzed in \citet{flury22}, hence a lower dust attenuation and a lower metallicity (see \Citealt{calabro21} for the relation between attenuation and metallicity), which translates into an increased capability to form channels through the ISM for the escape of ionizing radiation. Indeed, the normalization offset almost disappears when considering a subset from the LzLCS with a median UV slope similar to the NIRSpec galaxies ($\beta_{\rm median}$ $\simeq -2.1$).
A minor part of the offset is also due to our sample probing slightly lower stellar masses (by $\sim 0.3$ dex) than the sample of \citet{flury22}.

We also find that the slope of our correlation is consistent with the \fesc $\propto$ \sigmaSFR$^{0.4}$ relation predicted by the \citet{naidu20} model. 
We find a lower normalization of our relation by $\sim 0.3$ dex compared to their fiducial model with $a=1.6$, possibly suggesting a lower value of this parameter. For completeness, we also report in Fig. \ref{Fig:sigmaSFR_fesc}-\textit{top} the model with $a=0.8$, which is not ruled out in their work (as shown in their Fig. $6$), and whose prediction is much closer to our best-fit relation. However, their \fesc vs \sigmaSFR relation relies on measurements by \citet{steidel18}, which are based on a sample of star-forming galaxies at redshift $z\sim3$. The difference may thus originate from the different redshifts probed by our works, hence from the effects discussed above.  

A big caveat to the above analysis is that the two quantities on the x and y axis are not completely independent: \sigmaSFR correlates with the size, while \fesc is strongly dependent on the size by construction. While this is unavoidable, we also remark that, the \fesc estimation is based also on other galaxy properties and not just on the $r_e$.
However, in order to provide a more independent measurement, we also compare the estimated values of \fesc to the sSFRs, which do not directly depend on the physical extension of the galaxies (Fig. \ref{Fig:sigmaSFR_fesc}-\textit{bottom}). 
We find that the sSFR is still correlated to \fesc, although with a shallower slope and lower significance compared to the same relation with \sigmaSFR. Also in this case the slope of our best-fit relation is similar to what is found for the LzLCS galaxy sample at $z\sim0.3$, suggesting that the level of star-formation activity in a galaxy (regardless of the normalization method) has an impact on the escape of ionizing radiation. 
Overall, we are aware that all these quantities (i.e., \fesc, \sigmaSFR, sSFR, size, UV slope, and O32) are all interrelated somehow and not fully independent, as many of them depend on the same physical processes. This tells us that the best candidates as LyC leaking galaxies have rather peculiar and similar properties, being characterized by blue UV slopes, compact sizes, high O32, and also enhanced star-formation activity in the form of \sigmaSFR and sSFR.

\subsection{Connecting star-formation and outflows}\label{discussion:outflows} 

We have seen in the previous section that \sigmaSFR is tightly related to the ionizing properties of galaxies and their ability to spread ionizing photons in their surrounding medium. The underlying physical mechanism through which \sigmaSFR impacts on \fesc is that galaxies with more compact and enhanced star-formation activity facilitate the launch of galaxy scale outflows through stellar feedback, which includes radiation from young stars, stellar winds, and supernova explosions. Indeed, several studies have found higher gas outflow velocities and mass loading factors in galaxies with higher \sigmaSFR \citep{calabro22a,llerena23}, corroborating the close relation between \sigmaSFR and outflows. From the theoretical point of view, \citet{sharma16} and \citet{sharma17} predict that outflows are ubiquitous in galaxies with \sigmaSFR $\geq 10$ \msun/yr/kpc$^2$. 
The final connection between outflows and \fesc is then supported by a variety of works from the literature, both observational and theoretical \citep{heckman11,amorin12,borthakur14,sharma17,hogarth20}, according to which outflows can clear pathways in the surrounding neutral gas, favoring the escape of ionizing photons. 
Recently, \citet{amorin24} found a clear correlation between the velocity width of the \xOIII\ and \Ha\ broad-line wings and \fesc for a sample of $20$ Lyman continuum emitters at $z\sim0.3$, with the stronger emitters also having the highest values of \sigmaSFR.
In this subsection, we investigate spectral outflow signatures in our sample and study how they relate to the host galaxy properties. We also test the impact of outflows in an indirect way through the effect that they may have on the dust attenuation. 

\subsubsection{Exploring spectral outflow signatures in our sample}\label{subsection:outflow_signatures_individual}

\begin{figure*}[ht!]
    \centering
    \includegraphics[angle=0,width=0.92\linewidth,trim={0cm 0.5cm 10.8cm 0cm},clip]{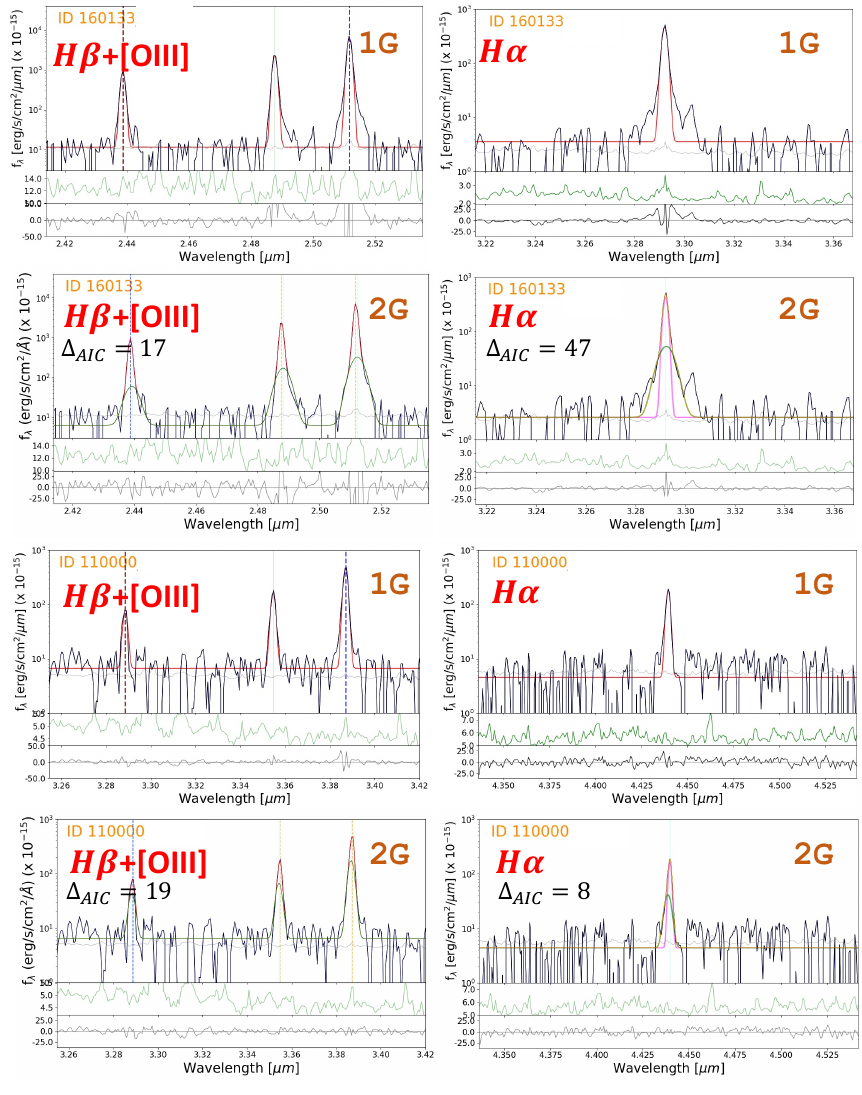}
    \vspace{-0.1cm}
    \caption{Diagrams showing the single Gaussian (1G) and double Gaussian (2G, narrow + broad) best fits to the \xOIII+\Hb and \Ha lines, as indicated in each panel label, for the two galaxies (ID GLASS 160133 and 110000) in which we detect a significant outflow signature in both emission lines (upper and lower four panels, respectively). In each panel, the upper part shows the observed spectrum with the best fit line in red (red (green) line for the narrow (broad) component in the double Gaussian fit), the middle part represents the error spectrum, while the bottom part is the residual after subtracting the best fit result. The spectra are in the observed frame wavelength. 
    }\label{Fig:outflow_individual}
    \vspace{-0.3cm}
\end{figure*}

To study the presence of outflows of ionized gas, we use the GLASS high resolution grating spectra and perform a single Gaussian and a double-component (narrow+broad) Gaussian fit to the \Ha and \xOIII\ lines (when available), which are the lines with the highest S/N in this redshift range, thus the most suited for exploring low S/N outflow features. 
In detail, \Ha was fitted alone, as the contribution from \xNII\ can be neglected at first order, while \xOIII\ was fitted together with \Hb and \xOIIItwo\ as explained in Section \ref{sec:line_measurements}. In the double Gaussian fit, we add a second broad component, allowing the intrinsic velocity width $\sigma$ of the narrow component to vary between $10$ and $250$ km/s, and $\sigma$ of the broad component from $100$ to $800$ km/s, which is the limit expected for star-forming driven outflows \citep{calabro22a}. We constrain the velocity shift between the two Gaussian centroids to $<300$ km/s, of the same order of the spectral resolution. We do not set any constraints on the flux ratio of the two components. In case of the \Hb + \OIII\ triplet, a broad component is added to all the three lines. 
Finally, we use the Akaike Information Criterion (AIC) criterion \citep{akaike74} to decide if the double Gaussian provides a significantly better fit to the line profile compared to a single Gaussian fit. In particular, we consider the double component as the final solution if the difference (AIC$_{\rm single}$ $-$ AIC$_\text{\rm double}$) $=$ $\Delta_{\rm AIC}$ is $>10$ \citep{burnham04,bosch19}, and if both the narrow and broad best-fit Gaussians are detected with a flux S/N $>3$. The above threshold ensures that the double component model is $100$ times more probable than the single component model to minimize the information loss. 


With the above criteria, we find that a broad component is required to fit \OIII\ and \Ha of two galaxies: ID 160133 and ID 110000 (Fig. \ref{Fig:outflow_individual}). Both galaxies have $\Delta_{\rm AIC}$ $> 15$ in \OIII (the highest S/N line), indicating a very strong case for the double component model. For the galaxy ID 110000, \Ha has a $\Delta_{\rm AIC}$ slightly lower than $10$ ($\sim 8$), but the double Gaussian is still highly favored ($50$ times more probable than a single component). 
We primarily interpret these components as an outflow, with $\sigma_{broad} =$ $250 \pm 4$ km/s and $200 \pm 10$ km/s, respectively. We measure a broad to narrow line flux ratio of $0.16$ ($0.3$) and a broad to narrow $\sigma$ ratio of $3.4$ ($2.8$). These galaxies lie at redshifts $4.02$ and $5.76$, respectively, and both have a modest $\log$ \sigmaSFR /(\msun yr$^{-1}$ kpc$^{-2}$) of $0.6 \pm 0.2$ ($0.4 \pm 0.2$), comparable to the average galaxy population at their redshifts, and $\log$ sSFR /(yr$^{-1}$) of $-7.6 \pm 0.2$ ($-8.5 \pm 0.3$). These systems are highlighted as big darkred stars in Fig. \ref{Fig:sigmaSFR_fesc}. The indirect estimation of \fesc for these two galaxies yields a value of $\simeq 0.11$ ($\simeq 0.04$), which is slightly above (slightly below) the median \fesc of galaxies at similar \sigmaSFR. We also note that these outflow velocities would correspond to strong leakers (\fesc $\geq 4$-$10 \%$) if we assume the $\sigma_{broad}$-\fesc correlation of \citet{amorin24}. 
Intriguingly, both galaxies with outflow signatures show low surface brigthness clumpy structures around the main cores from high-resolution JWST images. This makes the scenario of gas accretion an interesting, alternative possibility that cannot be excluded with our spectroscopic data. 

To better understand these results, we compare them to those obtained with a comparable dataset by \citet{carniani23}, which includes $52$ galaxies with R$=2700$ spectra from the JWST-JADES program at redshifts, stellar masses, and SFRs similar to our work. They detect broad additional components in \Ha or \OIII\ in a fraction of galaxies going from $15\%$ to $40\%$, which they interpret as evidence of outflows. They find that the outflow incidence increases with SFR but is rather stable (at the level of $20$-$30 \%$) as a function of \mstar and sSFR. 
Limiting the comparison to the redshift range between $4$ and $\sim 6$ (where most of their galaxies lie), and assuming the simple interpretation (also favored by their work) of an outflow scenario, we derive in GLASS an outflow incidence of $\sim20\%^{32\%}_{9\%}$ ($2$ out of $11$ galaxies with detected \OIII), with uncertainties derived following \citet{gehrels86}. This result is comparable to \citet{carniani23}. We also notice that our flux and $\sigma$ ratios between the narrow and broad Gaussian components fall within their observed range. All this suggests that we are probably probing the same physical phenomenon. 
Interestingly, we do not observe instead any outflow signatures in the $6$ GLASS sources at $z>6$ (upper limit of $26\%$). In this case, we cannot make comparisons as this redshift range is poorly sample by the other work. Within our sample, the number of high redshift sources with high resolution spectra is so low that the outflow incidence at $z>6$ is still compatible with that derived at lower redshift. However, we can exclude from this analysis that outflows become more important at higher redshifts. 

Furthermore, we do not find significant evidence of outflows in individual galaxies among the CEERS sample observed at medium resolution (R$\simeq1000$). To understand whether this is expected, and to test the effect of resolution on the outflow detectability, we perform a set of Monte Carlo simulations.
In detail, we simulate spectra at different redshifts injecting an emission line at a specific wavelength ($0.53 \mu m$ rest frame) assuming different line fluxes to produce a variety of S/N of the emission feature (from $3$ to $\sim 60$). We also set a broad to narrow intrinsic flux ratio in the range between $0.1$ and $0.7$, and different $\sigma$ ratios ranging $2$ to $12$, with $\sigma_{narrow,intrinsic}$ fixed to $50$ km/s. We vary the intrinsic broad line centroid within a resolution element. 
We then perform a single Gaussian and double Gaussian (narrow + broad) fit as done for the observed data, imposing the same requirements for taking the double component as the final best fit. Running these simulations ($100$ for each configuration) we find that the recovery rate is of the order of $2 \%$ or less for the typical range of outflow parameters expected from our works and from previous findings. This rate is not strongly dependent on the specific assumptions, except for the most extreme cases such as width and flux ratios $>=5$ and $>=0.5$, respectively. Therefore, the lack of significant outflow signatures in the CEERS sample is likely due to their lower spectral resolution. 

Increasing the S/N of the emission lines, for example through spectral stacking, can slightly increase the outflow recovery rate at median resolution and the possibility of detecting faint broad components in case of high outflow velocities. We thus stack the entire dataset selected in this work with H- or M-grating observations. We first downgrade by convolution the GLASS high resolution spectra to R $=1000$, convert the spectra to rest-frame using the spectroscopic redshifts determined in Section \ref{sec:line_measurements}, resample them to a common wavelength grid, and normalize them to the median flux estimated around the specific lines analyzed. Then, we apply for each pixel a median stacking with $5\sigma$ clipping, and finally perform the line measurements as done for individual galaxies. 
As shown by our simulations, this stacking approach tends to broaden a line systematically (regardless of the S/N or intrinsic line width) by $\sim50$ km/s, as due to the impact of the spectral resolution on the redshift estimation uncertainty when converting to rest-frame. 

Stacking the spectra around the \xOIII\ line, which is the highest S/N line available for the largest number of galaxies in our sample, and fitting the \Hb+\OIII\ triplet with a single and double Gaussian as described before, we find that the double component model is not strongly favored according to the previous condition ($\Delta_{\rm AIC} = 3$), thus there is no significant evidence of outflow.
Given the lower S/N, we are not able to appreciate significant differences in the fit or to detect any outflow signatures if we divide the sample in two or more bins of redshift, sSFR, or \sigmaSFR. 

To conclude, we find outflow incidence rates in individual galaxies observed by GLASS that are consistent to those obtained by similar works in the same redshift range. As highlighted by Carniani et al. (2023), the relatively low incidence rate is not an evidence of a minor role of outflows, as it could be related to the outflow geometry. For example, an incidence rate of $\sim 20\%$ can be obtained if outflows have a biconical morphology with an opening angle of $\sim 37 \deg$. 
This also suggests that, in order to robustly constrain the role of outflows in the \sigmaSFR - \fesc correlation, hence to confirm in a direct way the outflow-driven LyC photon escape scenario at the EoR, we need to study how the outflow fraction varies as a function of \sigmaSFR and \fesc, which requires a better statistics than that available in this analysis. This scientific question will be addressed in the future with a larger sample of galaxies with deep, high resolution NIRSpec data.

\subsubsection{Investigating the attenuation and the outflow incidence of extremely star-forming galaxies}\label{subsection:dust_clearing} 

\begin{figure}[t!]
    \centering
    \includegraphics[angle=0,width=1\linewidth,trim={0cm 0cm 0cm 0cm},clip]{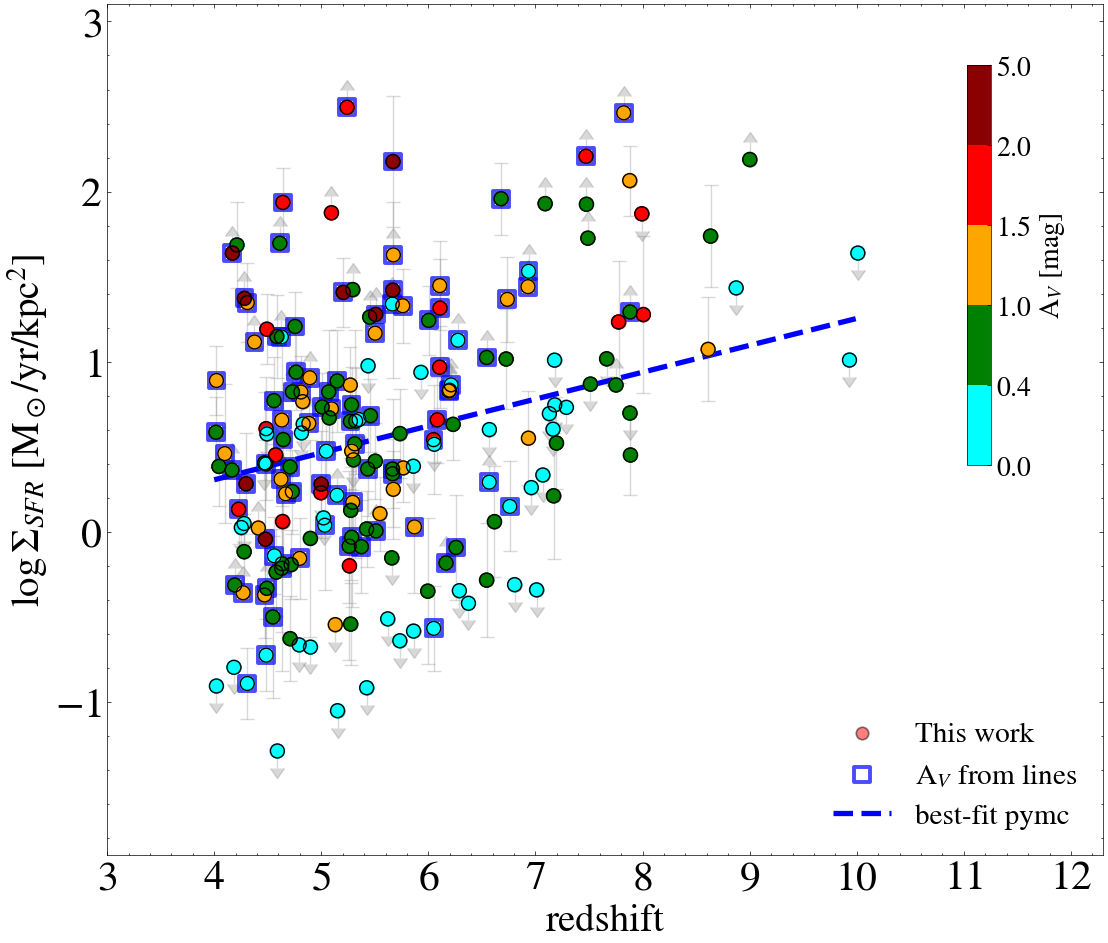}
    \vspace{-0.1cm}
    \caption{Redshift evolution of \sigmaSFR (analog to Fig. \ref{Fig:redshift_evolution}), color coded by the attenuation A$_V$, estimated as described in the text. Galaxies with A$_V$ derived from the emission lines are encapsulated within bigger blue squares.
    }\label{Fig:sigmaSFR_z_AV_coded}
    \vspace{-0.3cm}
\end{figure}

Approaching the epoch of cosmic dawn, galaxies tend to have denser ISM and more concentrated star-formation activity, as shown in the previous sections. For electron densities exceeding $\sim 200$ cm$^{-3}$, the radiation pressure from young stars starts to dominate over the thermal pressure sustained by supernovae explosions \citep{ferrara24}. Assuming the $(1+z)^p$ redshift evolution of $n_e$, we expect this transition to occur for the average galaxy population in a redshift range between $3$ and $7$, depending on the value of the exponent $p$ between $1$ and $2$ (see \Citealt{isobe23}). 
In these conditions of gas density and pressure, supernova explosions rapidly lose energy due to radiative losses, and the energy is converted into sound waves through shocks on a short timescale (see, e.g., \Citealt{martizzi15}, \Citealt{gatto15}, \Citealt{kim15}, \Citealt{walch15}, \Citealt{fielding17}), making supernova-driven outflows highly inefficient in early galaxies. Some theoretical models predict that in compact galaxies with extreme star-formation activity, a super-Eddington regime could be established, according to which radiation driven outflows can dramatically decrease the dust optical depth by pushing it to larger radii in a few Myr \citep{ferrara24}. This drastic event is expected when the sSFR exceeds $\sim 10^{-7.6}$ yr$^{-1}$. 
The consequence of this process is the formation of a system that has negligible dust attenuation, hence a very blue SED. This model was introduced to explain the excess of UV-bright and blue galaxies (compared to theoretical expectations) discovered during the first JWST observing campaigns \citep[e.g.,][]{naidu22,castellano22}, however it is still unclear whether it is the right physical explanation to the observational findings, and when exactly this proposed mechanism might start to dominate. 

To test this scenario we analyze the outflow incidence rate in galaxies with extremely high sSFR. First, we find that the fraction of galaxies in super Eddington regime (i.e., sSFR $> 10^{-7.6}$ yr$^{-1}$) in all the redshift range explored in this work ($4<z<10$) is consistent with the model predictions by \citet{ferrara24}, increasing from $\sim 15 \%$ at redshift $4$ to $\sim 40\%$ at redshift $10$. 
Among these galaxies, if we consider only those $5$ observed by GLASS at high resolution, we find an outflow incidence rate of $1$/$5$ ($\sim 20 \%$), or $1$/$3$ if we limit the analysis to systems with a compact morphology, that is, with an effective radius of $150$ pc or less. In both cases, these fractions are compatible with the incidence rate obtained from the full sample. 

\begin{figure}[t!]
    \centering
    \includegraphics[angle=0,width=1.00\linewidth,trim={0cm 0cm 0.cm 0cm},clip]{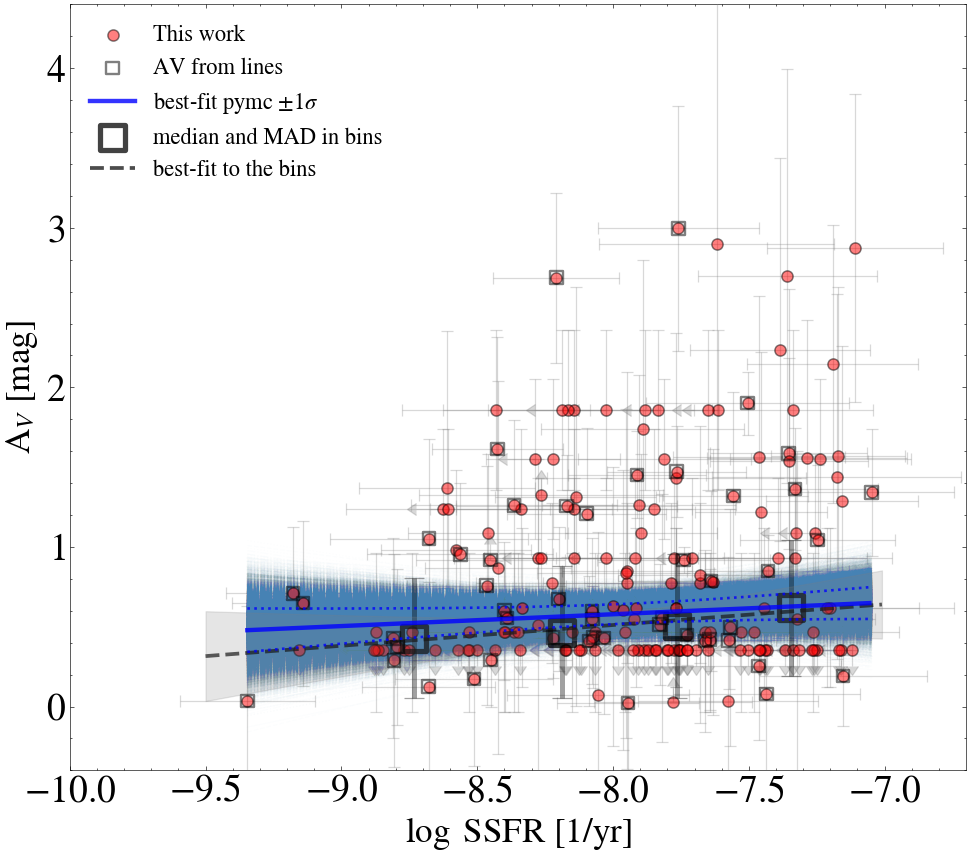}
    \vspace{-0.1cm}
    \includegraphics[angle=0,width=1.00\linewidth,trim={0cm 0cm 0cm 0cm},clip]{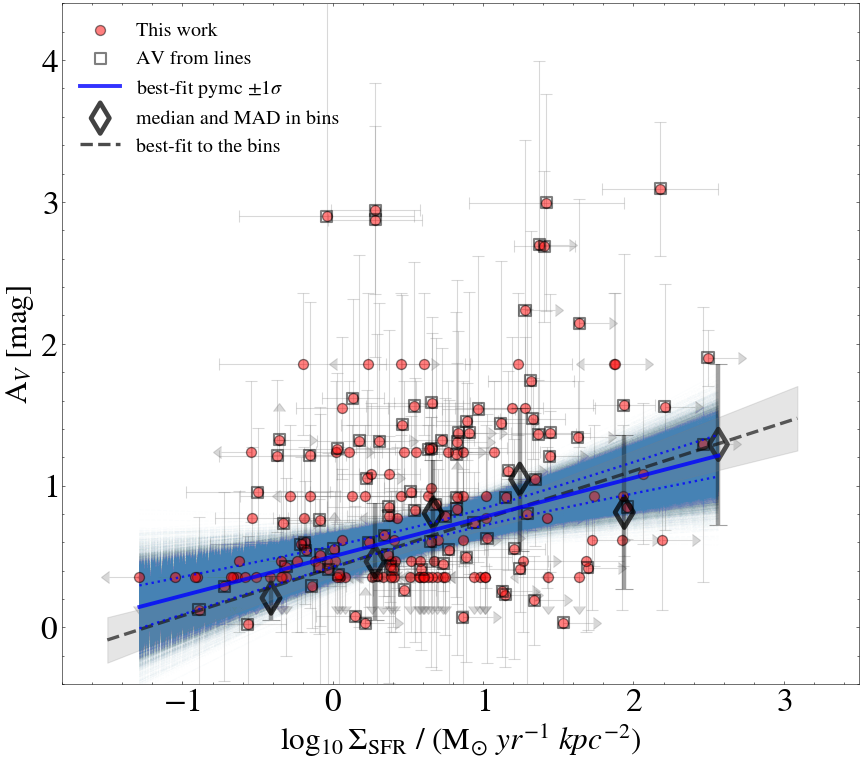}
    \vspace{-0.2cm}
    \caption{Diagrams comparing A$_V$ to sSFR and \sigmaSFR (\textit{top} and \textit{bottom} panels, respectively) for our selected galaxies in the redshift range $4<z<10$. The lines, shaded areas, and markers are the same as in Fig. \ref{Fig:redshift_evolution}. 
    }\label{Fig:sigmaSFR_AV}
    \vspace{-0.03cm}
\end{figure}

An alternative method to test the effect of star-formation on the dust content of galaxies, including the dust clearing scenario, is to analyze how the galaxy attenuation A$_V$ is related to the star-forming properties. 
To this aim, in Fig. \ref{Fig:sigmaSFR_z_AV_coded}, we color code the points in the $z$-\sigmaSFR diagram as a function of A$_V$. We can see that galaxies at $z>7$ have a large variety of A$_V$: while some galaxies have upper limits on A$_V$ consistent with $0$, some others have higher values that can reach A$_V = 2$ magnitudes. Therefore, we do not see evidence for a sudden drop of A$_V$ with redshift down to the very low values (A$_V$ $\simeq 10^{-2}$-$10^{-3}$) expected by the dust-clearing scenario, nor that the global galaxy population at $z>7$ is dominated by systems with A$_V \simeq 0$, even if the majority of this high-$z$ subset also has sSFRs largely exceeding the super-Eddington conditions defined above. 

We can better understand this result by plotting A$_V$ as a function of sSFR for the entire redshift range (top panel of Fig. \ref{Fig:sigmaSFR_AV}). We can see that galaxies with higher sSFR show a large variety of A$_V$. Globally, the median A$_V$ slightly increases with sSFR but it is consistent also with a flat relation ($m = 0.07 \pm 0.09$).  
In the bottom panel of Fig. \ref{Fig:sigmaSFR_AV} we analyze instead the relation between A$_V$ and \sigmaSFR. Here we find a positive correlation between the two quantities, with a best-fit slope of $0.28 \pm 0.07$ with the Bayesian method, indicating that galaxies with more concentrated star-formation (i.e., higher \sigmaSFR) have higher dust attenuations on average.
This positive correlation is similar to what is found in lower redshift studies \citep[e.g.,][]{reddy15,tomicic17}, and it reflects the fact that galaxies accumulate more dust in their ISM as a consequence of their increased (and more compact) star-formation activity. 
Dividing our sample in two redshift bins (lower and higher than $6$), we find no significant differences in the global trends seen in both panels of Fig. \ref{Fig:sigmaSFR_AV}. 

In conclusion, we can hypothesize several different explanations to our results. 
First, dust clearing radiatively driven outflows (in super Eddington conditions) have been proposed to explain the excess of bright and very blue galaxies at $z \geq 10$. We should note that the two NIRSpec galaxies at $z\simeq 10$ in our sample are both dust poor, consistent with this scenario.
We are focusing instead on a cosmic epoch where the feedback from supernovae is likely contributing to the acceleration of winds, as incorporated in several models and simulations \citep[e.g.,][]{sharma17}. Therefore, supernovae feedback remains the most plausible explanation for the outflows observed in NIRSpec galaxies at $z<6$. 
However, we should also be aware of the possible lesser role played by supernovae as we approach the reionization epoch, following the argument explained above. Galaxies at cosmic dawn are also expected to be younger than analogs at lower redshifts, with consequently shorter times available to supernovae to impact the surrounding ISM in a stable way. 

Secondly, the fraction of halo gas entrained by the outflowing shell might be conspicuous at the level of $10 \%$ or more. Consequently, the gas outflow velocities can be relatively low (see equation 15 in \Citealt{ferrara24}), to the level of $50$-$100$ km/s that are not easily detectable even in high resolution spectral mode. Another alternative explanation is that we are observing the galaxies in an intermediate phase of dust accumulation, possibly coincident with a period of gas accretion and dust creation by the first star formation episodes, that has not yet reached the conditions of dust opacity for developing a sudden blow-out event. In this particular phase, whose duration is not well known, the galaxies might still retain most of their gas and dust content, showing attenuations significantly higher than $0$ while keeping a low-level outflowing activity.

\section{Summary}

We summarize the main findings of this paper as follows. We analyze a spectroscopic sample of galaxies in the redshift range from $4$ to $10$ observed with JWST-NIRSpec by the CEERS and GLASS surveys. Using the Balmer lines and UV-based effective radii, we study the evolution of the SFR and \sigmaSFR over this cosmic epoch. 
\begin{enumerate}
\item We find that \sigmaSFR increases mildly in this redshift range (best-fit slope $=0.16 \pm 0.06$), rising by a factor of $2$ from $z=4$ to $10$. This trend is consistent with the predictions from several semi-analytic models of galaxy evolution, such as the Santa Cruz models \citep{somerville15}, and those by \citet{ferrara24} and \citet{naidu20}. 
\item A star-forming `Main Sequence' relation holds out to redshift $\sim 10$, with a best-fit slope of $\sim0.7$, consistent with the relations obtained for star-forming galaxies at lower redshifts.
\item We derive for the first time at $z > 4$ the \sigmaSFR `Main Sequence', finding a very mild increase (consistent with a flat relation) of \sigmaSFR with stellar mass. The slope of the best-fit relation is in agreement with observations at $0<z<2$, while the normalization is about $\sim2.5$ dex higher (at $z\sim10$) compared to the local Universe, resulting from the strong evolution of the average \sigmaSFR in star-forming galaxies during the last $13$ Billion years.
\item We show that \sigmaSFR is tightly related to the ionizing source properties, probed by the O32 index. This corroborates recent studies proposing \sigmaSFR as a key quantity responsible for the higher ionization parameter observed in high redshift galaxies \citep{reddy23}.
\item We find correlations of \sigmaSFR and sSFR with the escape fraction, estimated in an indirect way using the prescription of \citet{mascia23} calibrated on direct measurements of \fesc from the low-redshift Lyman continuum survey \citep{flury22}. The slopes of both correlations are in agreement with observations of low-redshift LyC leakers and with those predicted by the models of \citet{naidu20}. This suggests that \sigmaSFR and sSFR play an important role in the escape of ionizing radiation.  
\item We detect outflow signatures in \Ha and \xOIII\ lines for $\sim 20 \%$ of the galaxies at $z<6$ observed in H-grating NIRSpec mode, in agreement with the outflow incidence found in galaxies at similar redshifts by \citet{carniani23}. This relatively low incidence might be related to the outflow geometry. In particular, a biconical outflow morphology with an opening angle of $\sim 37 \deg$ is able to explain the observed fraction. A larger sample is instead necessary to study the outflow incidence at varying \sigmaSFR and \fesc, hence to have a more direct confirmation of the SFR - outflow connection and of the outflow-driven LyC photon escape scenario at the EoR.
\item We find a positive correlation between A$_V$ and \sigmaSFR, and a flat trend as a function of sSFR, indicating that there is no evidence of a drop of A$_V$ in extremely star-forming galaxies between $z=4$ and $10$. This might be at odds with a dust-clearing outflow scenario being in place at $z\geq 10$. 
\end{enumerate}

\noindent
Observing a larger number of galaxies at high resolution and testing the correlation between \sigmaSFR and outflow velocity predicted by supernovae feedback models can help us in the future to further constrain the properties and physical origin of the outflows detected both in our and in similar works at $z \sim 4$ to $10$, and further clarify the relation between star-formation and outflows in the epoch of reionization. Enlarging the sample of spectroscopically confirmed galaxies at $z>10$ is also essential to test possible transformations of the ISM properties occurring at this cosmic epoch.

\begin{acknowledgements}
We thank the referee for the detailed and constructive comments. AC acknowledges support from the INAF Large Grant for Extragalactic Surveys with JWST. We acknowledge support from the PRIN 2022 MUR project 2022CB3PJ3 - First Light And Galaxy aSsembly (FLAGS) funded by the European Union – Next Generation EU. We thank Stefano Carniani for useful discussion on the main topic of the paper. 
\end{acknowledgements}

{}

\end{document}